%% file: main.tex
\documentclass[letterpaper,twocolumn,10pt]{article}
\usepackage{usenix-2020-09}

\usepackage[noend]{algpseudocode}
\usepackage[ruled,vlined,linesnumbered,noend]{algorithm2e}
\usepackage{colortbl}
\usepackage{bbm}

\input{mydefs}

\begin{document}

\title{\mymethod: Differentially Private Trajectory Synthesis by Adaptive Markov Models}

\author{
{\rm Haiming Wang\textsuperscript{1}}
\ \ \
{\rm Zhikun Zhang\textsuperscript{2}\thanks{Zhikun Zhang is the corresponding author.}}
\ \ \ \ \
{\rm Tianhao Wang\textsuperscript{3}}
\ \ \ 
{\rm Shibo He\textsuperscript{1}}
\\
{\rm Michael Backes\textsuperscript{2}}
\ \ \
{\rm Jiming Chen\textsuperscript{1}}
\ \ \
{\rm Yang Zhang\textsuperscript{2}}
\\ 
\\
{\rm \textsuperscript{1}\textit{Zhejiang University}} \ \ \
\\
{\rm \textsuperscript{2}\textit{CISPA Helmholtz Center for Information Security}} \ \ \
\\
{\rm \textsuperscript{3}\textit{University of Virginia}} \ \ \
}

\maketitle
\pagestyle{plain}

\input{0_abstract}
\input{1_introduction}

\input{2_background}
\input{3_problem}

\input{4_overview}

\input{5_details}

\input{6_evaluation}

\input{7_discussion}
\input{8_related}

\input{9_conclusion}

{
    \footnotesize
    \bibliographystyle{plain}
    \bibliography{cite}
}

\appendix

\input{algorithm_analysis}

\input{app_method}

\input{app_more.tex}

\end{document}

%% file: mydefs.tex
\usepackage{booktabs,siunitx,multirow,multicol,tabularx}
\usepackage{epsfig,endnotes,xspace,url,diagbox}
\usepackage{amsmath,amsthm,amsfonts}
\usepackage{enumerate}
\usepackage{varioref}
\usepackage{graphicx}
\usepackage[font=footnotesize]{subfig}
\usepackage{lipsum}
\usepackage{tikz}
\usepackage{float}
\usepackage{epstopdf}
\usepackage{xcolor}
\usepackage{hyperref}
\usepackage{comment}
\usepackage[utf8]{inputenc}
\usepackage{mathtools}
\usepackage{wrapfig}
\usepackage{xspace}
\usepackage{algorithm2e}

\usepackage{cases}
\usepackage{stfloats}
\usepackage{fancyhdr}
\usepackage{bbm}
\usepackage{colortbl}
\usepackage{enumitem}

\definecolor{mygray}{gray}{.9}

\hypersetup{
  colorlinks,
  linkcolor={blue!70!green},
  citecolor={green!70!blue},
  urlcolor={orange!70!red}
}

\newcommand{\lemmaref}[1]{\hyperref[#1]{Lemma~\ref*{#1}}}

\usepackage[hang,flushmargin]{footmisc}
\usepackage{titlesec}
\titlespacing*{\section}{0pt}{*3}{4pt}
\titlespacing*{\subsection}{0pt}{*2}{3pt}
\titlespacing*{\subsubsection}{0pt}{*1}{2pt}
\let\oldbibliography\thebibliography
\renewcommand{\thebibliography}[1]{%
  \oldbibliography{#1}%
  \setlength{\itemsep}{4pt}%
}

\newtheorem{theorem}{Theorem}
\newtheorem{lemma}[theorem]{Lemma}

\newtheorem{definition}{Definition}

\newcommand{\mypara}[1]{\noindent\textbf{#1.} \xspace}

\renewcommand{\Pr}[1]{\ensuremath{\mathsf{Pr}\left[#1\right]}\xspace}

\newcommand{\mymethod}{\ensuremath{\mathsf{PrivTrace}}\xspace}
\newcommand{\adatrace}{\ensuremath{\mathsf{AdaTrace}}\xspace}
\newcommand{\dpt}{\ensuremath{\mathsf{DPT}}\xspace}

\renewcommand{\AA}{\mathcal{A}\xspace}
\newcommand{\myexp}[1]{\ensuremath{e^{#1}}\xspace}
\newcommand{\Lapp}[1]{\ensuremath{\mathcal{L}\left(#1\right)}\xspace}

\newcommand{\dset}{\ensuremath{\mathcal{D}}\xspace}
\newcommand{\stateset}{\ensuremath{\Sigma}\xspace}
\newcommand{\markovstate}{\ensuremath{\sigma}\xspace}
\newcommand{\stateseq}{\ensuremath{T}\xspace}
\newcommand{\seqstate}{\ensuremath{\lambda}\xspace}

\newcommand{\jsd}{\ensuremath{\mathsf{JSD}}\xspace}
\newcommand{\are}{\ensuremath{\mathsf{ARE}}\xspace}

\newcommand{\realtrajectory}{\ensuremath{S}\xspace}
\newcommand{\statetrajectory}{\ensuremath{T}\xspace}
\newcommand{\cellinmodel}{\ensuremath{C}\xspace}

\newcommand{\levelonedipar}{\ensuremath{K}\xspace}
\newcommand{\leveltwodipar}{\ensuremath{\kappa}\xspace}
\newcommand{\levelonedensity}{\ensuremath{d}\xspace}
\newcommand{\firstorderm}{\ensuremath{M_{1}}\xspace}
\newcommand{\secondorderm}{\ensuremath{M_{2}}\xspace}
\newcommand{\tripdis}{\ensuremath{\{t_{ij}\}}\xspace}
\newcommand{\densityvec}{\ensuremath{\eta}\xspace}
\newcommand{\marquery}{\ensuremath{\gamma}\xspace}
\newcommand{\tripset}{\ensuremath{\phi}\xspace}
\newcommand{\tripnum}{\ensuremath{t}\xspace}

\newcommand{\countend}{\ensuremath{q}\xspace}
\newcommand{\countstart}{\ensuremath{b}\xspace}

\newcommand{\statenumber}{\ensuremath{m}\xspace}

\newcommand{\transitioncount}{\ensuremath{N}\xspace}
\newcommand{\harlength}{\ensuremath{l}\xspace}

\newcommand{\marorder}{\ensuremath{k}\xspace}

\newcommand{\transition}{\ensuremath{P}\xspace}

\newcommand{\modelselectionpar}{\ensuremath{\theta}}
\newcommand{\smallvalueinexp}{\ensuremath{\varphi}\xspace}

\newcommand{\markovmodel}{Markov chain model\xspace}
\newcommand{\markovmodels}{Markov chain models\xspace}
\newcommand{\randomwalk}{random walk\xspace}

\newcommand{\appendixref}[1]{\hyperref[#1]{Appendix~\ref*{#1}}}

\usepackage{etoolbox}
\makeatletter
\patchcmd{\hyper@makecurrent}{%
    \ifx\Hy@param\Hy@chapterstring
        \let\Hy@param\Hy@chapapp
    \fi
}{%
    \iftoggle{inappendix}{%
        \@checkappendixparam{chapter}%
        \@checkappendixparam{section}%
        \@checkappendixparam{subsection}%
        \@checkappendixparam{subsubsection}%
        \@checkappendixparam{paragraph}%
        \@checkappendixparam{subparagraph}%
    }{}%
}{}{\errmessage{failed to patch}}

\newcommand*{\@checkappendixparam}[1]{%
    \def\@checkappendixparamtmp{#1}%
    \ifx\Hy@param\@checkappendixparamtmp
        \let\Hy@param\Hy@appendixstring
    \fi
}
\makeatletter

\newtoggle{inappendix}
\togglefalse{inappendix}

\apptocmd{\appendix}{\toggletrue{inappendix}}{}{\errmessage{failed to patch}}

%% file: 0_abstract.tex
\begin{abstract}
Publishing trajectory data (individual's movement information) is very useful, but it also raises privacy concerns.
To handle the privacy concern, in this paper, we apply differential privacy, the standard technique for data privacy, together with \markovmodel, to generate synthetic trajectories.
We notice that existing studies all use \markovmodel and thus propose a framework to analyze the usage of the \markovmodel in this problem.
Based on the analysis, we come up with an effective algorithm \mymethod that uses the first-order and second-order Markov model adaptively.
We evaluate \mymethod and existing methods on synthetic and real-world datasets to demonstrate the superiority of our method.
\end{abstract}

%% file: 1_introduction.tex
\section{Introduction}
\label{sec:intro}

Trajectory data analysis plays an important role in tasks related to the social benefit such as urban planning and intelligent transportation~\cite{wang2020deep,atluri2018spatio}.
The trajectory data used usually comes from users who carry mobile devices to record their locations.
However, the sensitive nature of the trajectory data also gives rise to privacy risks when being shared. 
Recent research demonstrated effective privacy attacks despite aggregation or anonymization, such as trajectory reconstruction~\cite{gursoy2018utility}, de-anonymization~\cite{chang2018revealing}, and membership inference~\cite{pyrgelis2017knock,CZWBHZ21}.
Without a sufficient privacy protection technique, trajectory data analysis is not possible since users will refuse to share their data.

A promising technique to overcome the privacy concern is \textit{differential privacy} (DP)~\cite{dwork2006calibrating,ZWLHC18,DZBLJCC21,WCZSCLLJ21}, which has become the golden standard in the privacy community.
Intuitively, on a given dataset, DP defines a random transformation process (algorithm) so that the output of the transformation contains some ``noise'' that can ``cover'' the existence of any possible record in the dataset.
DP can be used by both companies and government agencies.  With DP protection, hopefully, more users will be willing to share their data. From the government side, agencies can also share more data for the public good.

Most of the previous studies on differentially private trajectory data analysis focus on designing tailored algorithms for specific tasks, such as
community discovering~\cite{xu2018dp}, participatory sensing~\cite{li2017achieving}, and recommendation~\cite{wei2019differential}.
Our paper focuses on a more general approach, where we publish a {\it synthetic trajectory dataset} that shares similar properties with the original one while satisfying DP.
While publishing synthetic datasets would not work better than directly publishing aggregate statistics~\cite{stadler2021synthetic}, this paradigm can easily enable any (even unseen) down-stream data analysis tasks without modifying existing algorithms and has been adopted in, e.g., the 2020 US Census data publication~\cite{CensusBureau} and a dataset for population flow analysis~\cite{iwata2019neural}. 
There are several recent studies focusing on generating data satisfying DP, and they are typically composed of two steps: {\it model learning} and {\it generation}~\cite{zhang2017privbayes,zhang2022privsyn,mckenna2019graphical}.
In the scenario of trajectory data, because the information is typically continuous, existing works~\cite{gursoy2018utility,he2015dpt} also need to first {\it discretize} the data upfront.

\mypara{Existing Work}
The first work to generate synthetic trajectories with DP is \dpt by  He et al.~\cite{he2015dpt}.
It uses a set of uniform grids with different granularities to discretize the space and establishes multiple differentially private prefix trees~\cite{chen2012differentially} (each with a different grid) to model transitions; then generates data by random walk on the prefix trees.
More recently, Gursoy et al.~\cite{gursoy2018utility} proposed \adatrace.  
It discretizes the geographical space into two kinds of grids using different granularities. 
Then a first-order \markovmodel and other three important features are learned based on the discretization results.
Then the data is generated by random walk on the first-order \markovmodel with help of extracted features.
The drawback of \adatrace is that information contained in the first-order \markovmodel is not enough to generate data of high quality.

\mypara{Our Contributions} 
Existing approaches either get only the first-order \markovmodel, which fails to obtain enough transition information, or get a high-order \markovmodel, which introduces excessive noise due to DP.
Our idea is to reach a middle ground between \dpt and \adatrace.
To this end, we employ only the first and second-order \markovmodel, for three reasons.
First, previous studies have shown that the second-order \markovmodel can achieve promising accuracy for next step prediction~\cite{song2006evaluating,gambs2012next,karimzadeh2018pedestrians}.
Second, \markovmodel of higher order is too space and time-consuming.
A \marorder -order \markovmodel with \statenumber states needs $\statenumber^{\marorder + 1}$ transition probability values.
When \statenumber is 300, and \marorder is 3, this requires 30GB of storage.
Third, to satisfy DP, building more models will introduce more noise.
Although more information is extracted, the amount of noise also grows.  
As a result, after some specific order of the Markov chain model, the overall information quality will decrease.  
In practice, we observe the threshold is between the second-order and third-order.

To generate high-quality trajectories, we use both the first-order model and the second-order model in our random-walk-based generation process.
Every time to predict a state as the next step, we select between the first-order model and the second-order model.
The selection principle contains two aspects of considerations: the effect of noise on different models and the confidence ability of models.

The standard \markovmodel does not have the information about where the trajectory starts and ends. 
We add virtual start and end in every trajectory to record this information.
However, due to some actions to bound the sensitivity, the recorded information is biased.
To reduce the bias, we propose an optimization-based method to estimate the trip distribution.
The optimization problem is built based on the observation that a trajectory only contributes one to the count related to the virtual start and end.
The experimental results show that our estimation method is effective, especially when the privacy budget is small.

We conduct empirical experiments on both synthetic and real-world trajectory datasets to compare \mymethod with the state-of-the-art.
\mymethod consistently outperforms the existing methods for a variety of metrics.
We further conduct comprehensive ablation studies to illustrate the effectiveness of the three main components of \mymethod.
One limitation of our approach is that the accuracy of the generated long trajectories is worse than that of the short trajectories.  
This is due to the fact that long trajectories are more complicated and more challenging to generate correctly.

To summarize, the main contributions of this paper are two-fold:

\begin{itemize}[leftmargin=*]
\setlength\itemsep{-0.25em}
    
    \item We propose a new trajectory data synthesis method \mymethod.  
    Its key insight is to exploit both the first-order and second-order \markovmodels. 
    \mymethod is built with a new method to choose between the first-order and second-order transition information for next-step prediction, and a new method to estimate the trip distribution from the first-order \markovmodel without consuming extra privacy budget.

    \item We conduct extensive experiments on both synthetic and real-world trajectory datasets to validate the effectiveness of \mymethod.  
    Our code is open-sourced at \url{https://github.com/DpTrace/PrivTrace}.
\end{itemize}

\vspace{-1mm}

\mypara{Roadmap} 
In \autoref{sec:background}, we give the definition of the problem and present background knowledge of DP. 
Then we show the framework of trajectory data generation \autoref{sec:problem}.
Following this framework, we provide the design of our method in \autoref{sec:overview}.
The experimental results are presented in \autoref{sec:experiments} with discussions in \autoref{sec:discussion}.

Finally, we discuss related work in \autoref{sec:relatedwork} and provide concluding remarks in \autoref{sec:conclusion}.

%% file: 2_background.tex
\section{Preliminaries}
\label{sec:background}

\subsection{Problem Definition}
\label{subsec:problem_definition}

In this paper, we consider a dataset consisting of a set of \textit{trajectories}.  
Each trajectory is composed of a \textit{sequence of points}.
We are interested in the following question:
Given a sensitive trajectory dataset $\dset_o$, how to generate a synthetic trajectory dataset $\dset_s$ that shares similar properties with $\dset_o$ while satisfying DP.
Generating the synthetic trajectory dataset facilitates down-stream data analysis tasks without modifications.

Following prior work~\cite{gursoy2018utility}, we use four statistical metrics to measure the similarity between $\dset_s$ and $\dset_o$: length distribution, diameter distribution, trajectory density, and transition pattern.
The {trajectory length} is the total distance of a trajectory, which can be used to study the commute distance of people.
The {trajectory diameter} measures the largest Euclidean distance between any two points in a trajectory, which gives information about the range of individual's activities.
The \textit{length distribution} and \textit{diameter distribution} capture the frequency of different trajectory lengths and trajectory diameters in a trajectory dataset, respectively.
We use Jensen-Shannon divergence (\jsd)~\cite{lin1991divergence} to measure the similarity of distributions between $\dset_o$ and $\dset_s$.
A smaller \jsd means the generated dataset $\dset_s$ is more similar to the original $\dset_o$.

The \textit{trajectory density} calculates the number of trajectories that pass through a given area on the map, which can be a good indicator for urban planning, such as estimating the flow of traffic in a specific area.
The \textit{transition pattern} captures the frequency of transiting from one place to another, which can help solve problems like next location prediction.
We use the average relative error (\are) of different randomly generated queries of trajectory density and transition pattern to measure their similarity between $\dset_o$ and $\dset_s$.
A smaller \are implies $\dset_s$ is more similar to $\dset_o$.

\subsection{Markov Chain Model}
\label{subsec:markov_model}
To generate a synthetic dataset, a commonly used method is the \markovmodel.  A \markovmodel is a stochastic model describing a sequence of possible events in which the probability of each event depends only on the state attained in the previous events.
It is frequently used to analyze sequential data such as location trajectories~\cite{goh2012online,lv2016big,gambs2012next,bashir2007object,cheng2016improved} and natural language~\cite{morwal2012named,fine1998hierarchical,juang1991hidden,tokuda2013speech}.
Formally, the \markovmodel is defined as follows:
\vspace{-1mm}
\begin{definition}[Markov Chain Model]
    Given a finite set of discrete states $\stateset=\{\markovstate_{1}, \markovstate_{2}, \markovstate_{3}, \ldots, \markovstate_{k}\}$, a sequence $\stateseq=(\seqstate_{1}, \seqstate_{2}, \seqstate_{3}, \ldots \seqstate_{s})$ is said to follow a $\marorder$th-order Markov process if $\marorder \leq i \leq s - 1,  \forall \seqstate \in \Sigma$
    $$
    \begin{array}{crl}
        \Pr{\seqstate_{i+1} = \markovstate_{j} | \seqstate_{i} \ldots \seqstate_{1}} 
        = \Pr{\seqstate_{i+1} = \markovstate_{j} | \seqstate_{i} \ldots \seqstate_{i - \marorder}} \\
    \end{array}
    $$
where $\Pr{\seqstate_{i+1} | \seqstate_{i} \ldots \seqstate_{1}}$ is called the transition probability from $\seqstate_{i} \ldots \seqstate_{1}$ to $\seqstate_{i+1}$.

\end{definition}

Given a dataset \dset, if $r$ is a subsequence of any $\stateseq \in \dset$, the empirical transition probability $\Pr{\markovstate|r}$ is defined as
\begin{align}
    \Pr{\markovstate|r} & = \frac{\sum\limits_{\forall \stateseq \in \dset} \transitioncount_{\stateseq}(r\sigma)}{\sum\limits_{\forall \stateseq \in \dset} \sum\limits_{\forall x \in \Sigma}\transitioncount_{\stateseq}(rx)}=\frac{\transitioncount_{\dset}(r\sigma)}{\sum\limits_{\forall x \in \Sigma}\transitioncount_{\dset}(rx)}
    \label{equ:probability_calculation}
\end{align}
Here $rx$ is a sequence where $r$ is followed by $x$.
$\transitioncount_{\stateseq}(rx)$ is the total number of occurrences of $rx$ in $\stateseq$, and we denote $\sum\limits_{\forall \stateseq \in \dset}\transitioncount_{\stateseq}(rx)$ as $\transitioncount_{\dset}(rx)$.

The $\marorder$th-order \markovmodel is the \markovmodel that can provide \Pr{\markovstate|r} for any state $\markovstate$ and any subsequence $r$ with length $\marorder$.
By calculating all $\transitioncount_{\dset}(rx)$ for all possible length-$\marorder$ subsequence $r$ and all $x \in \Sigma$, we can learn a $\marorder$th-order \markovmodel from the dataset.

\subsection{Differential Privacy}
\begin{definition}[$\epsilon$-Differential Privacy] \label{def:non-pure-dp}
	An algorithm $\AA$ satisfies $\epsilon$-differential privacy ($\epsilon$-DP), where $\epsilon>0$,
	if and only if for any two neighboring datasets $\dset$ and $\dset'$, we have
	\begin{equation*}
    	\forall{O\subseteq\! \mathit{Range}(\AA)}:\; \Pr{\AA(\dset)\in O} \leq e^{\epsilon}\, \Pr{\AA(\dset')\in O},
    	\label{eq:npdp}
	\end{equation*}
	where $\mathit{Range}(\AA)$ denotes the set of all possible outputs of the algorithm $\AA$.
\end{definition}

In this paper, we consider two datasets $\dset$ and $\dset'$ to be neighbors, denoted as $\dset \simeq \dset'$ if and only if $\dset = \dset^{'} + \realtrajectory$ or $\dset^{'} = \dset + \realtrajectory$, where $\dset + \realtrajectory$ denotes the datasets resulted from adding one trajectory $\realtrajectory$ to the dataset $\dset$.

\mypara{Laplacian Mechanism}
The Laplace mechanism computes a function $f$ on input dataset $\dset$ while satisfying $\epsilon$-DP, by adding to $f(\dset)$ a random noise.
The magnitude of the noise depends on $\mathsf{GS}_f$, the \emph{global $L_1$ sensitivity} of $f$, defined as,
\[
\mathsf{GS}_f = \max\limits_{\dset\simeq \dset'} ||f(\dset) - f(\dset')||_1
\] 
When $f$ outputs a single element, such a mechanism $\AA$ is given below:
\vspace{-1.3mm}
$$
\begin{array}{crl}
& \AA_f(\dset) & =f(\dset) + \Lapp{\frac{\mathsf{GS}_f}{\epsilon}}
\end{array}
$$
where $\Lapp{\beta}$ denotes a random variable sampled from the Laplace distribution with scale parameter $\beta$ such that $\Pr{\Lapp{\beta}=x} = \frac{1}{2\beta} \myexp{-|x|/\beta}$.

When $f$ outputs a vector, $\AA$ adds independent samples of $\Lapp{\mathsf{GS}_f/\epsilon}$ to each element of the vector.
The variance of each such sample is $2 \mathsf{GS}_f^2/\epsilon^2$.

\subsection{Composition Properties of DP}
\label{subsec:composition}
The following composition properties are commonly used for building complex differentially private algorithms from simpler subroutines. 

\mypara{Sequential Composition}
Combining multiple subroutines that satisfy DP for $\epsilon_1, \cdots,\epsilon_k$ results in a mechanism that satisfies $\epsilon$-DP for $\epsilon=\sum_{i} \epsilon_i$.

\mypara{Post-processing}
Given an $\epsilon$-DP algorithm $\AA$, releasing $g(\AA(\dset))$ for any $g$ still satisfies $\epsilon$-DP.  That is, post-processing an output of a differentially private algorithm does not incur any additional privacy concerns.

%% file: 3_problem.tex
\begin{table*}[!t]
    \caption{Summary of existing methods on different steps.}
	\centering
	\begin{tabular}{c|c|c|c}
		\toprule
		\backslashbox{Method}{Step}	& Discretization & Model Learning & Generation \\ 
		\midrule
		
		\adatrace~\cite{gursoy2018utility} & Adaptive partition & First-order \markovmodel & Markov sampling + auxiliary info \\
		\dpt~\cite{he2015dpt} & Reference system & Multiple Prefix Trees & Markov sampling \\
		\mymethod & Adaptive partition & First-/second-order \markovmodel & Random walk \\
		\bottomrule
	\end{tabular}
	\label{tbl:framework_summary}
\end{table*}

\section{Existing Solutions}
\label{sec:problem}

Before diving into the descriptions of existing methods that use \markovmodel to generate synthetic trajectories, we first present a general recipe that we observe in all these solutions.

\subsection{A Framework for Markov-based Trajectory Synthesis}
\label{subsec:framework}

To generate synthetic trajectories using the \markovmodel, there are three major components: 
Geographical space discretization, model learning, and trajectory generation.
We integrate all the components in a general framework:
\begin{itemize}[leftmargin=*]
\setlength\itemsep{-0.25em}
    \item \mypara{Discretization}
    The purpose of discretization is to create discrete states for the \markovmodel.
    We discretize the continuous geographical space into one or more \textit{grids} where each grid partitions the geographical space.
    After discretization, each area in the grid is regarded as a state in the \markovmodel.
    
    \item \mypara{Model Learning}
    Given a set of geographical states and a trajectory dataset, we need to learn some models that can capture the transition pattern of the trajectory dataset.  
    Here, the model can be a \markovmodel and other information (e.g., trip distribution) extracted from the dataset.
    To train the \markovmodel, we calculate all $\transitioncount_{\dset}$ needed in \autoref{equ:probability_calculation} and then add noise to achieve DP. 
    
    \item \mypara{Generation}
    After the model is learned, we can generate the synthetic trajectory (e.g., by random walk on the model).
    The trajectory generation component is a post-processing step in the context of DP, thus does not have an additional privacy concern.
\end{itemize}

\subsection{Existing Methods}
\label{subsec:existing_methods}

\autoref{tbl:framework_summary} summarizes these three phases of existing work.  
In what follows, we review these steps for previous studies.

\mypara{AdaTrace}
This method extracts the first-order \markovmodel, trip distribution, and length features from the sensitive dataset with DP, and then generates private trajectories according to these features. 
In discretization, \adatrace proposes to discretize the geographical space into grids twice with two different granularities.
It first uniformly discretizes the geographical space into a coarse-grained grid.
For the cells in the first grid with a large number of trajectories passing through, \adatrace further discretizes them into finer-grained grids.

In model learning, \adatrace learns a first-order \markovmodel using the first grid.  
To better capture the inherent patterns of the dataset, \adatrace also extracts the length distribution and trip distribution (count of pairs of starting and ending point in the trajectory).
Noise is added to achieve DP.

In generation, \adatrace first samples a starting state and an ending state from the trip distribution, and samples a trajectory length $|T|$ from the length distribution.
Then, \adatrace generates the trajectory from the starting state, and repeatedly generates the next state by random walking on \markovmodel.
After $|T|-1$ steps, the trajectory directly jumps to the ending state.
Finally, for each generated state, \adatrace uniformly samples a location point from the corresponding cell.
If the cell is further partitioned, the location point is sampled according to the density of the second-layer grid.

\mypara{DPT}
In order to capture more precise information, \dpt uses multiple uniform grids with different granularities.  
When discretizing trajectories, \dpt will choose the most coarse-grained grid for which there exists a transition for two consecutive points of the trajectory (i.e., using a more coarse-grained grid, the two locations in the trajectory will be in the same cell).  A trajectory will be transformed into multiple segments of transition, where transitions in the same segment are on the same grid.

Now for each grid, \dpt establishes a differentially private prefix tree~\cite{chen2012differentially} to model the transitions using the transition segments.  
A trajectory will be generated following the path on the prefix tree.
Note that there are also transitions to allow a state in one prefix tree to move to another prefix tree with different granularity.

%% file: 4_overview.tex
\section{Our Proposal}
\label{sec:overview}

Both \adatrace and \dpt adopt \markovmodel to capture the transition pattern of the trajectory dataset.
But they work in two extremes:
\adatrace mostly uses the first-order \markovmodel to reduce the amount of noise.
The drawback is that the first-order \markovmodel only captures a limited amount of information, and thus the result is not accurate.
On the other hand, \dpt uses high-order \markovmodel, but this leads to excessive noise being added, and also makes the result inaccurate.

Intuitively, there is a trade-off between the amount of information we can extract and the amount of noise we have to add due to the constraint of DP.  
Our insight is to work in the middle ground between the two extremes of \adatrace and \dpt, where we use the Markov chain model in a way that is neither too coarse-grained nor too fine-grained.  
In particular, we employ the first-order and second-order \markovmodel and select useful information in the two models for generation.
To this end, we propose \mymethod, which achieves a good trade-off between accuracy and noise.

\subsection{Method Overview}
\label{subsec:method_overview}

\begin{figure*}
    \centering
    \includegraphics[width=1\textwidth]{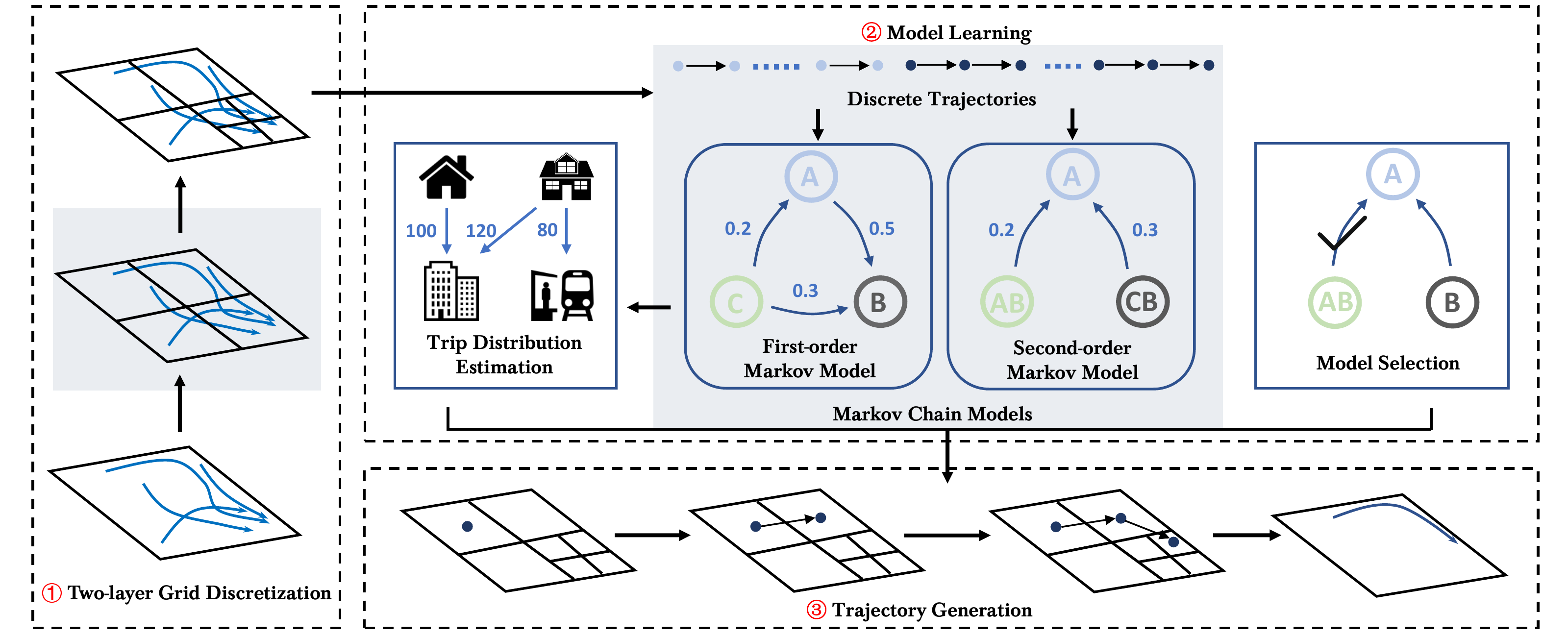}
    \vspace{-1mm}
    \caption{Method overview.
    \mymethod is composed of three parts: discretization, model learning, and trajectory generation.
    The discretization step first partitions the space into a coarse-grained uniform grid, and then the density of the cells with DP to determine which cells need to be expanded.
    For model learning, we learn both the first-order and second-order Markov chain models with differential privacy, and obtain the trip distribution from the two models.
    The trajectory generation process is a random-walk-based algorithm to generate a synthetic trajectory.
    We propose a method to select from the two models during the generation of synthetic trajectories.
    }
    \label{fig:method_overview}
\end{figure*}

\mymethod follows the general framework for generating synthetic trajectory data described in \autoref{sec:problem}, which consists of three major parts: geographical space discretization, model learning, and trajectory generation.

\mypara{Step 1: Discretization}
To better capture the transition information of the trajectories, we rely on the observation that the places with more trajectories passing through should be discretized in a finer-grained manner.
To this end, we use a density-sensitive two-layer discretization scheme.
The core idea is to first discretize the map into a coarser-grained uniform grid (or {first-layer grid}), and the cells in the first-layer grid with many trajectories passing through are further discretized into finer-grained (second-layer) grid.
The details of this step are in \autoref{subsec:grid}.

\mypara{Step 2: Model Learning}
Our model learning step contains two components: the Markov chain models learning and trip distribution estimation.

\textit{Markov Chain Models Learning.}
We first estimate the first-order and second-order models in a differentially private manner.
The details of this component are discussed in \autoref{subsec:markov_model_building}.

\textit{Trip Distribution Estimation.}
We estimate the trip distribution to obtain the frequency of the starting and ending points of the trajectories.
The \markovmodel already contains this, but it is biased due to a normalization step when training the \markovmodel.
To this end, we propose a method to obtain an accurate estimation of the distribution estimation.
The details of this component are referred to \autoref{subsec:optimization}.

\mypara{Step 3: Generation}
We use a random-walk-based method to generate trajectories, which starts at a random state and predict the next state by using either first-order or second-order \markovmodel.
Concretely, we first sample a pair of starting and ending states from the estimated trip distribution and \randomwalk with the two \markovmodels.
The details of this step are referred to \autoref{subsec:trajectory_generation}.

Note that we use two \markovmodels to predict the next step in the synthetic trajectory. In the prediction process, the first-order and second-order \markovmodels have different prediction abilities.
To take advantage of both two models, we propose two criteria to choose between them, which will be described in \autoref{subsec:model_selection}.

%% file: 5_details.tex
\subsection{Geographical Space Discretization}
\label{subsec:grid}

We borrow the idea of Qardaji et al.~\cite{dpgrid2013} to discretize the geographical space.
In particular, we first divide the geographical space into $\levelonedipar \times \levelonedipar$ cells (we call them the first-layer cells) of equal size. 
For each cell, we further calculate the number of trajectories that passes through this cell, and if the number is large, we further partition it.

To estimate the occurrences of trajectories for all first-layer cells, we use the Laplacian mechanism with parameter $\epsilon_1$.
Here one challenge is that the sensitivity is unbounded, since a trajectory can have an arbitrary number of occurrences in cells.

This introduces infinite DP noise.
To address this issue, we normalize the trajectory (the normalization method introduced in~\cite{gursoy2018utility}) to bound the sensitivity (proved in \appendixref{app:method_grid}).
For example, in \autoref{fig:method_overview}, we first discretize the space into four cells $C_1, C_2, C_3,$ and $C_4$.  
Consider a trajectory that occurs in $C_1$ and $C_2$, its total number of occurrences in all cells is 2.
After we normalize the occurrence of the trajectory in $C_1$ and $C_2$ by the total occurrence, the trajectory contribute to the trajectory occurrence in $C_1$ and $C_2$ by $1/2$ and $1/2$ instead of $1$ and $1$. 

If a cell has many trajectories passing over it, we expand it using a more fine-grained cell so that we can understand that area with more details.  
In \autoref{fig:method_overview}, step 1, the right bottom cell is expanded into four cells.

\mypara{Difference from \adatrace}
\adatrace~\cite{gursoy2018utility} adopts a similar two-layer discretization scheme as \mymethod; however, the use of two-layer grids is different.
In \adatrace, the cells in the first layer are used as states for \markovmodel, while the cells in the second layer are used for sampling in the generation phase.
On the other hand, in \mymethod, the cells from both the first-layer grid (if not divided again in the second layer) and the second-layer grid are used as states.
The advantage of using cells from all layers as states is that it can capture finer-grained transition information.

\subsection{Markov Chain Models Learning}
\label{subsec:markov_model_building}

After the two-layered discretization, the whole space is split into disjoint areas, denoted by $\{C_i\}$. 
To learn the Markov chain models, we calculate the counts of each possible transition $\transition$.
As our goal is to train the first-order and second-order Markov models, we only consider $\transition$ of length 2 or 3.
Note that for the purpose of sampling, there should be a virtual starting and an ending states in the Markov models.  To incorporate this, we augment each trajectory with a start and an end state.

Similar to the case of occurrence estimation in discretization, here the sensitivity of the transition counts is also unbounded (since a sequence can be long and lead to a large change in transition counts), and we also use a length normalization method to bound the sensitivity (by dividing the counts by the length of the trajectories) to $1$.

Concretely, denoting the occurrence of transition $\transition$ in the trajectory $\stateseq$ as $N_{\stateseq}(\transition)$, the transition count of $\transition$ in the \markovmodel is defined as the length-normalized value $\sum_{\forall \stateseq \in \dset} \frac{N_{\stateseq}(\transition)}{|\stateseq|}$.
Using such a definition, every trajectory contributes to the transition count of $\transition$ in the \markovmodel by $\frac{N_{\stateseq}(\transition)}{|\stateseq|}$, which is at most 1.
Therefore, the change of one trajectory will change the counts in the \markovmodel by at most 1
(see details in \autoref{pf:transition_model_sensitivity}).

We then add Laplace noise to the transition counts (we spend $\epsilon_2$ and $\epsilon_3$ for the length-2 and length-3 transitions, respectively) and use the noisy transition counts to build the Markov chain models.

Note that adding noise might make the transition count negative, which makes the sampling of trajectory infeasible.
To deal with this issue, we adopt the postprocessing method NormCut~\cite{wang2020locally,zhang2022privsyn} to handle the negative values.
We refer the readers to \autoref{subapp:negative_value} for the details of NormCut.

\subsection{Trip Distribution Estimation}
\label{subsec:optimization}
The normalization operations in~\autoref{subsec:markov_model_building} introduce bias to the transition counts (intuitively, we over-count short trajectories and under-count long trajectories because they have equal weights after normalization) used in the Markov models.
In this subsection, we propose a novel method to reduce bias.
The key idea is to leverage the spacial structure and the assumption that people tend to follow the shortest path when traveling.  
Therefore, for trajectories starting from location $i$ and ending at $j$, the normalization factor is the length of the shortest path from $i$ to $j$.
Building on this assumption, we estimate the distribution of trips between any pair of $i$ and $j$.

Specifically, denote $t_{ij}$ as the number of trajectories from $i$ to $j$, such that $\sum\limits_{i,j=1, i\neq j}^m t_{i,j}$ equals the number of trajectories in the original dataset.  From the Markov model, we know the normalized transition counts from the virtual starting point to any first location $i$, denoted by $\countstart_{i}$, and from any last location $j$ to the virtual ending point, denoted by $\countend_{j}$.  Moreover, we know the shortest paths from $i$ to $j$, denoted by $l_{ij}$.
Ideally, we have
\begin{align*}
    \countstart_i\simeq \sum\limits_{j=1}^m \frac{t_{ij}}{\harlength_{ij}}\mbox{ \;\; and \;\;} \countend_j \simeq \sum\limits_{i=1}^m \frac{t_{ij}}{\harlength_{ij}}
\end{align*}

which captures the intuition that the number of normalized trajectories starting from $i$ should equal the summation (over the ending location $j$) of all normalized trajectories starting from $i$, and similarly for the normalized trajectories ending at $j$.
Now we have $2m$ approximate equations with $m^2$ unknown variables (the $t_{ij}$'s).  
We cannot directly solve the unknowns.  
Instead, we build an optimization problem to estimate approximate values for them.
In particular, we use existing solvers to find $t_{ij}$'s that can minimize the following quantity:

\begin{align*}
&\min_{t_{i,j}}\quad \sum\limits_{i=1}^m \left(\sum\limits_{j=1}^m \frac{t_{ij}}{\harlength_{ij}} -\countstart_i \right)^2 + \sum\limits_{j=1}^m \left(\sum\limits_{i=1}^m \frac{t_{ij}}{\harlength_{ij}} - \countend_j \right)^2 \\
& \begin{array}{r@{\quad}r@{}l@{\quad}l}
s.t.&\sum\limits_{i,j=1}^m t_{i, j}& = |\dset|  \\
 &t_{i,j}&\geq0,  &i, j = 1, 2, 3, \ldots, m  \\
\end{array}
\end{align*}

More details of the optimization problem are given in \autoref{subapp:method_optimization}.

Given the trip distribution, next, we will then describe how to use it in the final generation process.
We empirically show in \autoref{exp:ablation_study} that using the trip distribution estimated by our method can achieve better accuracy than that of directly extracting from the original dataset, especially when the privacy budget is small.

\begin{algorithm}[!t]
\SetAlgoLined
\LinesNumbered
\KwIn{
Noisy first-order and second-order transition models $M_1$ and $M_2$, trip distribution $\{t_{ij}\}$;}
\KwOut{Synthesis trajectory Dataset $\dset_{s}$;}

Set $\dset_{s}$ as an empty set;

\For{$|\dset_{s}| \leq n_{syn}$}{
    \textbf{Step 1: Initialization}
    
    Set $T_{state}$ as an empty sequence of state;
    \label{In:initialization1}
    
    Sample state $\seqstate_{start}$ from the trip distribution $\{t_{ij}\}$;
    
    $\seqstate_{last} \leftarrow \seqstate_{start}$;
    $\seqstate_{now} \leftarrow \firstorderm(\seqstate_{start})$
    ;

    \label{In:initialization2}
    
    \textbf{Step 2: Random Walk}
    
    Add $\seqstate_{start}$ to $T_{state}$;
    
    \While{$\seqstate_{now}$ is not virtual end}{
        
        $T_{state} \leftarrow T_{state}\seqstate_{now}$;

        \mypara{Step 2-1: Model Selection}
       
        \eIf{$\transitioncount_{now,sum} < \modelselectionpar_1$ or $\transitioncount_{now,1} / \transitioncount_{now,2} \ge \modelselectionpar_2$}{\firstorderm is selected}{\secondorderm is selected}

       \mypara{Step 2-2: Model Prediction}
       
       \eIf{\secondorderm is selected}{
        $\seqstate_{next} \leftarrow \secondorderm(\seqstate_{now}, \seqstate_{last})$;
            \label{In:rw1}
        }
        {
        $\seqstate_{next} \leftarrow \firstorderm(\seqstate_{now})$;
        \label{In:rw2}
        }
        
        $\seqstate_{last} \leftarrow \seqstate_{now}$; $\seqstate_{now} \leftarrow \seqstate_{next}$;
    }
    
    \textbf{Step 3: Location Point Sampling}

    $T_{syn}\gets$ sampling a location for every state in $T_{state}$;
    \label{In:ls1}
    
    $\dset_{s} \leftarrow \dset_{s} \cup \{T_{syn}\}$;
}
\caption{Synthetic Trajectory Generation}
\label{alg:trajectory_generation}
\end{algorithm}

\subsection{Trajectory Generation}
\label{subsec:trajectory_generation}

Our trajectory generation algorithm relies on random walking on the first- or second-order Markov models.
The workflow of the trajectory generation algorithm is illustrated in \autoref{alg:trajectory_generation}.
It takes as input the first-order and second-order \markovmodels \firstorderm and \secondorderm, and the trip distribution \tripdis, and works in three steps as follows:

\mypara{Step 1: Initialization}    
We first set $T_{state}$ as an empty sequence of states.
Then a start-end state pair ($\seqstate_{start}$, $\seqstate_{end}$) is sampled from the trip distribution \tripdis.
The predicted real end state $\seqstate_{end}$ will not be used in the generation process.
We estimate it mainly to make the optimization problem in \autoref{subsec:optimization} more accurate.
We then set $\seqstate_{last}$ as the $\seqstate_{start}$ 
and $\seqstate_{now}$ as $\firstorderm(\seqstate_{start})$ (\autoref{In:initialization1}-\autoref{In:initialization2}).

\mypara{Step 2: Random Walk}
We first add $\seqstate_{now}$ to the end of $T_{state}$.
Then the next state prediction is conducted.
Before predicting the next step, we select a proper model to use.
The model selection method will be explained in \autoref{subsec:model_selection}.
Concretely, if \secondorderm is the chosen model, then the next state $\seqstate_{next}$ is $\secondorderm(\seqstate_{now}, \seqstate_{last})$ 
(\autoref{In:rw1}).
Otherwise, \firstorderm is the chosen model, $\seqstate_{next}$ is $\firstorderm(\seqstate_{now})$ 
(\autoref{In:rw2}).
$\firstorderm(\seqstate_{now})$ represents predicting $\seqstate_{next}$ only relying on the \textit{current state} $\seqstate_{now}$, and $\secondorderm(\seqstate_{now}, \seqstate_{last}))$ represents predicting $\seqstate_{next}$ relying on both the current state $\seqstate_{now}$ and the \textit{previous state} $\seqstate_{last}$.
After acquiring  $\seqstate_{next}$, we set $\seqstate_{last}$ as $\seqstate_{now}$.
Then value of $\seqstate_{next}$ is given to $\seqstate_{now}$.
The above process is repeated until $\seqstate_{now}$ is the virtual end.

\mypara{Step 3: Location Point Sampling}
After the random walk process, we obtain the discrete version trajectory $T_{state}$.
It is a state sequence.
We first set the synthetic trajectory sequence $T_{syn}$ as an empty sequence.
Then we sample locations for all states in $T_{state}$ and add all these locations into $T_{syn}$ 
(\autoref{In:ls1}).
Specifically, for every state, we sample a location from the geographical area corresponding to it uniformly.
$T_{syn}$ is the output synthetic trajectory.

\subsection{Markov Chain Models Selection}
\label{subsec:model_selection}

Given $\firstorderm$ and $\secondorderm$, one core question in random walk is how to choose between them in the trajectory generation process.
Without loss of generality, supposing we have already generated $i$ states $(\seqstate_{1}, \dots, \seqstate_{i})$, we need to determine whether to predict the \textit{next state} $\seqstate_{i+1}$ relying on the first-order model $\firstorderm(\seqstate_{i})$ or relying on the second-order model $\secondorderm(\seqstate_{i-1}, \seqstate_{i})$.

\mypara{Selection Rationales}
To determine which model to select, we need to consider two important factors: One is the noise error introduced by the Laplacian noise, and the other is the prediction confidence of the two models.
Intuitively, if the noise has a significant impact on both models, we select the model with less noise.
Otherwise, we select the model with higher prediction confidence.

\mypara{Selection Principles}
The model selection is mainly based on the count of transitions in $\marquery^{1}(\dset)$. 
Given the current state $\seqstate_{i}$, the count of transitions from $\seqstate_{i}$ to any possible state is denoted as $\transitioncount_{i}=\{\transitioncount_{\dset}(\seqstate_{i}\seqstate_{j}), \forall \seqstate_{j} \in \Sigma\}$.
For the sake of brevity, we denote the largest and the second largest count value in $\transitioncount_{i}$ as $\transitioncount_{i, 1}$ and $\transitioncount_{i, 2}$, and the sum of all counts as $\transitioncount_{i, sum}$.
We analyze the advantages and disadvantages of \firstorderm and \secondorderm from two aspects:
\begin{enumerate}[leftmargin=*]
\setlength\itemsep{-0.25em}
    \item \mypara{Count vs. Noise}
    If $\transitioncount_{i, sum}$ is smaller than a threshold $\modelselectionpar_{1}$, we use \firstorderm as the model to predict the next step; otherwise we will consider the next principle.
    
    The reason for this rule is that noise has a larger impact on the second-order model than the first-order model since the true counts in the second-order model are always smaller than or equal to those in the first-order model.
    Thus, the first-order model has a larger signal-to-noise ratio, and will more likely to bring better performance when the counts are small.

    \item \mypara{Existence of Dominant State}
    If the counts are not too small, we will compare $\frac{\transitioncount_{i, 1}}{\transitioncount_{i, 2}}$ to a threshold $\modelselectionpar_{2}$.  If $\frac{\transitioncount_{i, 1}}{\transitioncount_{i, 2}} \ge \modelselectionpar_{2}$, we use \firstorderm as the model to predict the next step, otherwise we will use \secondorderm.
    
    Here, we use the term $\frac{\transitioncount_{i, 1}}{\transitioncount_{i, 2}}$ as an indicator of whether there is a dominant state.
    If $\frac{\transitioncount_{i, 1}}{\transitioncount_{i, 2}} \ge \theta_2$, the state corresponds to $\transitioncount_{i, 1}$ is much more likely to be chosen as the next state than any other states in \firstorderm.

\end{enumerate}

In summary, when the state count is large enough ($N_{i,sum}\theta_1$) and the largest count ($N_{i,1}$) is not dominant, we prefer to choose \secondorderm; otherwise, we prefer to choose \firstorderm.

The values of $\theta_1$ and $\theta_2$ are hyperparameters, which we will discuss and verify in the evaluation (see \autoref{subapp:verify_model_selection_parameters}).

\subsection{Algorithm Analysis}
\label{subsec:algorithm_analysis}

\mypara{DP Guarantee}
Here we show that \mymethod theoretically satisfies $\epsilon$-DP.

\begin{theorem}
\mymethod satisfies $\epsilon$-differential privacy, where $\epsilon = \epsilon_1 + \epsilon_2 + \epsilon_3$.
\label{thm:dp_proof}
\end{theorem}

\noindent We refer the readers to \autoref{app:dp_proof_detials} for the detailed proof of \autoref{thm:dp_proof}.

\mypara{Computational Complexity Analysis}
Due to space limitation, we refer the readers to \autoref{app:complexity_analysis} for the detailed time and space complexity analysis of \mymethod, \adatrace, and \dpt.
We also empirically evaluate the time and space consumption of different methods.
We also discuss the practical time consumption in \autoref{sec:discussion}.

%% file: 6_evaluation.tex
\section{Evaluation}
\label{sec:experiments}

In this section, we first conduct an end-to-end experiment to illustrate the superiority of \mymethod over the state-of-the-arts.
We then conduct comprehensive ablation studies to illustrate the effectiveness of different components of \mymethod.
Finally, we compare the utility of the synthetic trajectories with different lengths.

\subsection{Experimental Setup}
\label{subsec:experimental_setup}

\mypara{Datasets}
We run experiments on one synthetic and two real-world trajectory datasets.
The statistics of the datasets are summarized in \autoref{table:dataset}.

\begin{itemize}[leftmargin=*]
\setlength\itemsep{-0.25em}
    \item \mypara{Brinkhoff~\cite{brinkhoff2002framework}}
    Brinkhoff is a popular network-based trajectory generator in the field of traffic research.
    Given the road network of a certain area, it can generate the trajectories residing in the roads.
    We use the road network data of California bay area to generate trajectories.
    
    \item \mypara{Taxi~\cite{moreira2013predicting}}
    It contains more than 1.7 million trajectories from 442 taxis in Porto, Portugal.
    We randomly select 200,000 trajectories to conduct our experiment.

    \item \mypara{Geolife~\cite{zheng2009mining}}
    This dataset is collected from volunteers recruited by Microsoft Research Asia carrying GPS devices.
    It contains 17,621 trajectories, with a total distance of more than 1.2 million kilometers and a total time of more than 48,000 hours.
    Most trajectories are in Beijing, China.
\end{itemize}

\begin{table}[!t]
    \centering
    \caption{Dataset Statistics.}
    \label{table:dataset}
    \setlength{\tabcolsep}{1.5mm}
    \begin{tabular}{c c c c c}
    \toprule
    \bf{Dataset} & \bf{Type}    & \bf{Area}    & \bf{Scale}\\  
    \midrule
    Brinkhoff    &  Synthetic   &   California & 30,000 \\   
    \rowcolor{mygray}
    Taxi         &  Real        &   Porto      &    200,000   \\
    Geolife      &  Real        &   Beijing    &   17,621    \\             
    \bottomrule
    \end{tabular}
\end{table}

\mypara{Metrics}
We adopt four metrics to measure the similarity between the original dataset $\dset_o$ and synthetic dataset $\dset_s$:
Length distribution, diameter distribution, trajectory density, and transition pattern.
Due to space limitation, we refer the readers to \autoref{subapp:evaluation_metrics} for the detailed description of these metrics.

Note that \adatrace also uses these metrics in their experiments; however, our parameter settings are different, which might lead to inconsistent values for the \adatrace method with those reported in their paper.
Our parameter settings aim for more fine-grained information.
For the length and diameter distribution metrics, \adatrace uses 20 bins to obtain the distribution while we use 50 bins.
For the trajectory density metric, \adatrace considers the query areas with a constant radius, while we consider random radii.
For the transition pattern metric, \adatrace uses a $6 \times 6$ uniform grid to obtain the transition patterns, while we use a $20 \times 20$ uniform grid.

\mypara{Competitors}
We compare \mymethod with \dpt and \adatrace discussed in \autoref{subsec:existing_methods}.
We use their open-sourced implementations to conduct our experiments, i.e., \adatrace~\cite{adatracecode} and \dpt~\cite{DPTcode}.
We also use the recommended parameters from their papers.
It is worth noting that, after carefully checking the code snippet of the length extraction process of \adatrace, we find that some steps do not satisfy differential privacy.
For a fair comparison, we modify the corresponding code snippets to make them differentially private and run the experiment using the modified code.
Due to space limitation, we refer the readers to \autoref{appsub:competitors} for the modification of the code.

\begin{figure*}[!htb]
    \centering
    \subfloat{\includegraphics[width=1\textwidth]{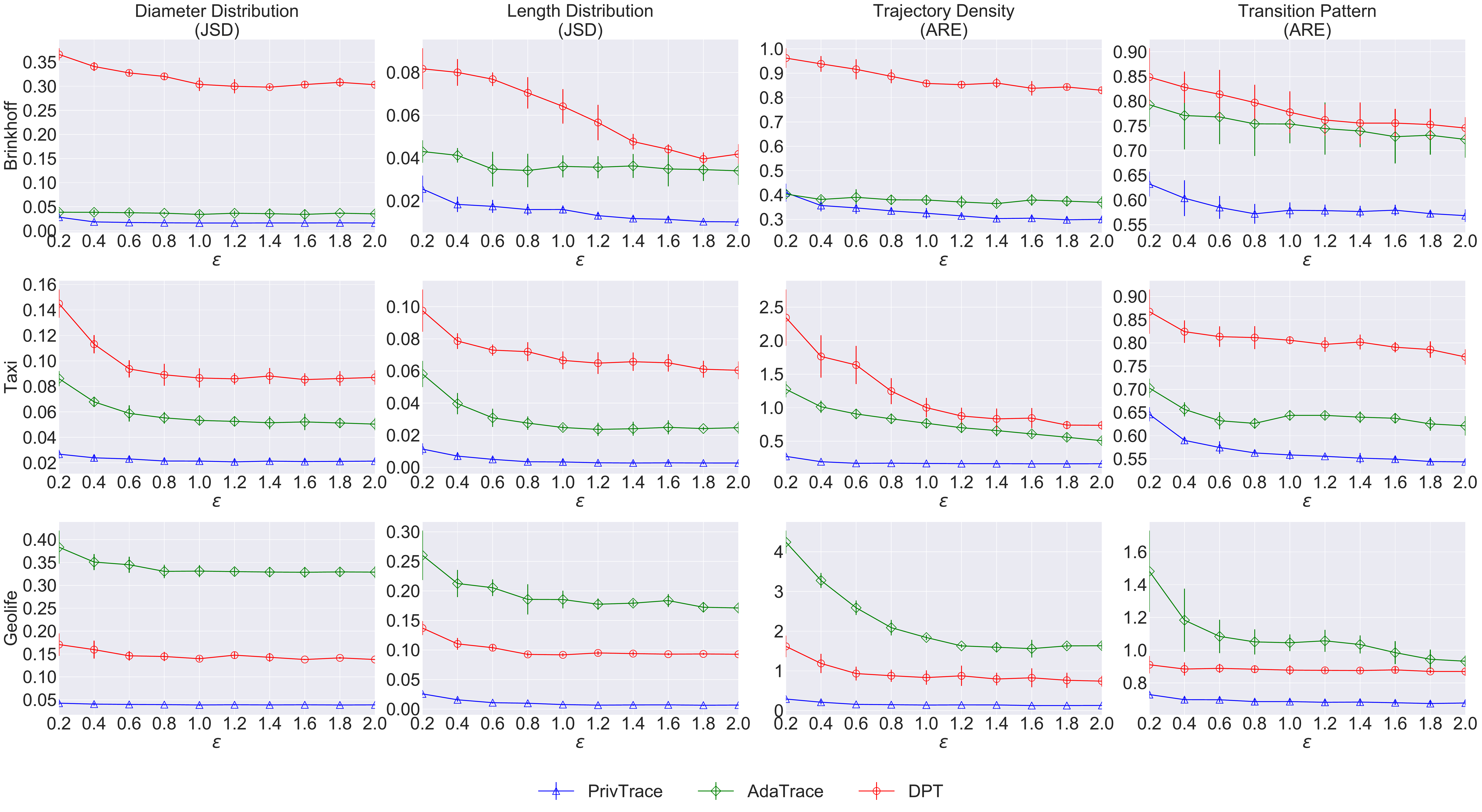}}
    \caption{End-to-end comparison of different trajectory synthesis methods.
    Rows and columns stand for different datasets and different metrics respectively.
    In each subfig, the $y$-axis stands for the error of the generated trajectories, and the $x$-axis stands for the privacy budget $\epsilon$.
    }
    \label{fig:comparison_private}
\end{figure*}

\mypara{Parameter Settings}
\mymethod has two groups of parameters: discretization parameters $\levelonedipar$, $\leveltwodipar$ in \autoref{subsec:grid} and model selection parameters $\theta_1$, $\theta_2$ in \autoref{subsec:model_selection}.

The choice of $\levelonedipar$ is related to whether the trajectories are uniformly or concentratedly distributed on the map and the number of trajectories.
A dataset with more concentratedly distributed trajectories should have fewer first-layer cells to reduce states which are unrelated to trajectories.
Therefore, $\levelonedipar$ is set as $(|\dset|/c)^{\frac{1}{2}}$, where $c$ is a parameter related to the distribution of trajectories.
We set $c$ as 5000, 1200, and 500 for Brinkhoff, Taxi, and Geolife, respectively.
The corresponding $\levelonedipar$ are 3, 13, 6.
One the other hand, $\leveltwodipar$ is calculated by $(\levelonedensity_{i} \times \levelonedipar \times pop / 2\times 10^7)^{\frac{1}{2}}$, where $pop$ is the population of the area where the trajectories are residing in.
We obtain the information of $pop$ from the Internet.
We verify the effectiveness of the parameter setting of $\levelonedipar$ and $\leveltwodipar$ in \autoref{subapp:verify_discretization_parameters}.

For $\theta_1$, the standard deviation of the noise added to the transition counts of each state is $(\frac{\sqrt{2}}{\epsilon_2}) \cdot m$, where $\frac{\sqrt{2}}{\epsilon_2}$ is the standard deviation of Laplace noise, $m$ is the total number of states.
Therefore, we set $\theta_1$ to $(\frac{\sqrt{2}}{\epsilon_2}) \cdot m$ to choose states with transition counts comparable to noise.
On the other hand, $\theta_2$ is used for choosing states with dominant transition states.
We empirically find that setting $\theta_2$ as 5 works well since we believe when the largest transition count is 5 times the second-largest transition count, the transition state with the largest count dominates other transition states.
We verify the effectiveness of these parameter settings in \autoref{subapp:verify_model_selection_parameters}.

For \adatrace and \dpt, we use their default parameter settings in experiments.

\mypara{Implementation}
We use different $\epsilon$ in our experiments, ranging from 0.2 to 2.0. 
We set $\epsilon_1 = 0.2\epsilon, \epsilon_2 = 0.4\epsilon$, and $\epsilon_3 = 0.4\epsilon$ (see \autoref{subapp:impact_budget_allocation} for the empirical results of the effectiveness of this privacy budget allocation).
Each experiment is repeated 10 times with mean and standard deviation reported.

We implement \mymethod with Python 3.6 and NumPy 1.19.1.
All experiments are run on an Intel E5-2680 server with 128 GB memory and Ubuntu 20.04 LTS system.

\subsection{End-to-end Evaluation}
\label{subsec:end_to_end_evaluation}

\begin{figure*}[!htpb]
    \centering
    
    \subfloat[$\mathsf{Real}$]{\includegraphics[width=0.24\textwidth,height = 0.14\textwidth]{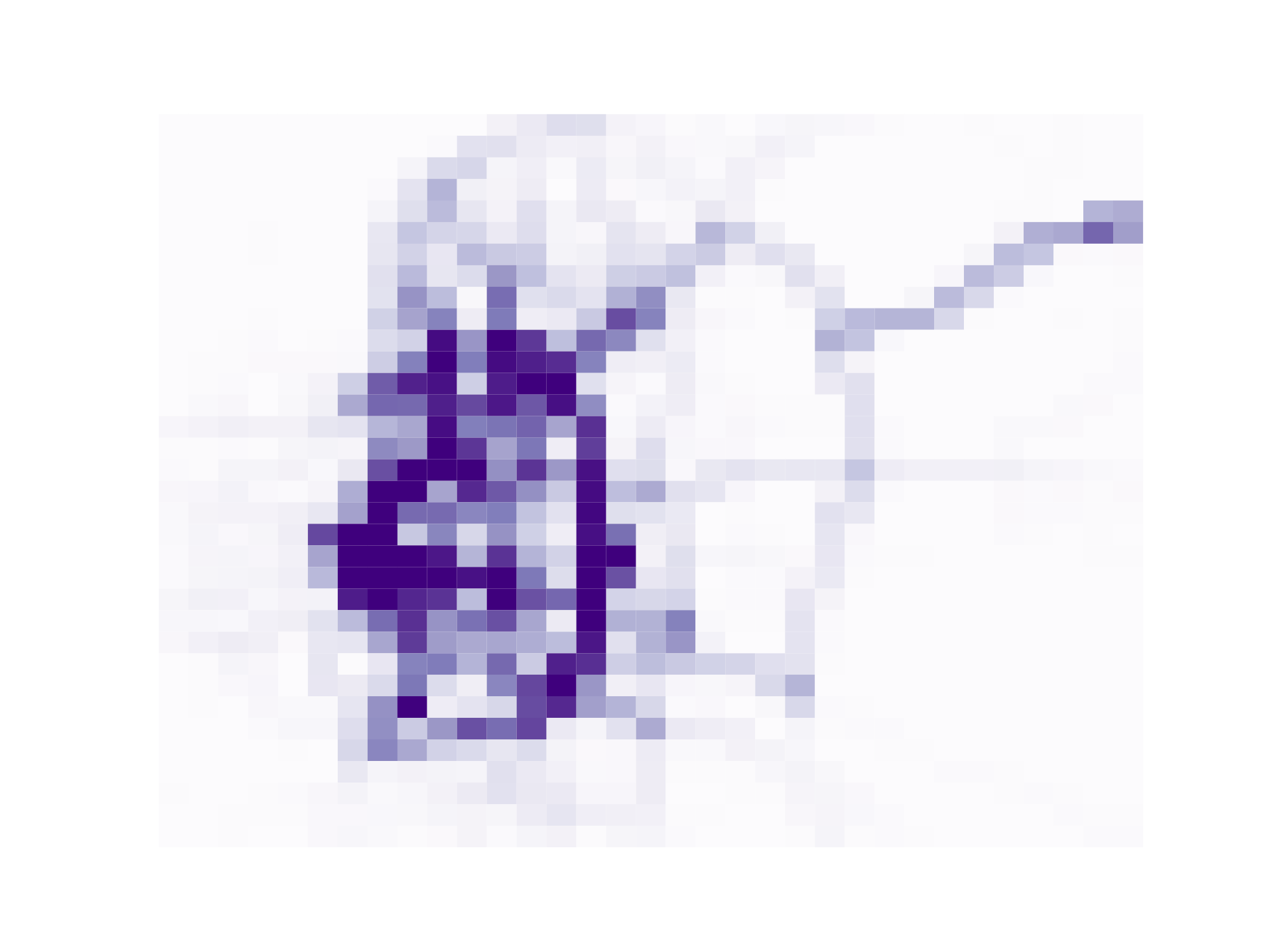}}
    \subfloat[\mymethod]{\includegraphics[width=0.24\textwidth,height = 0.14\textwidth]{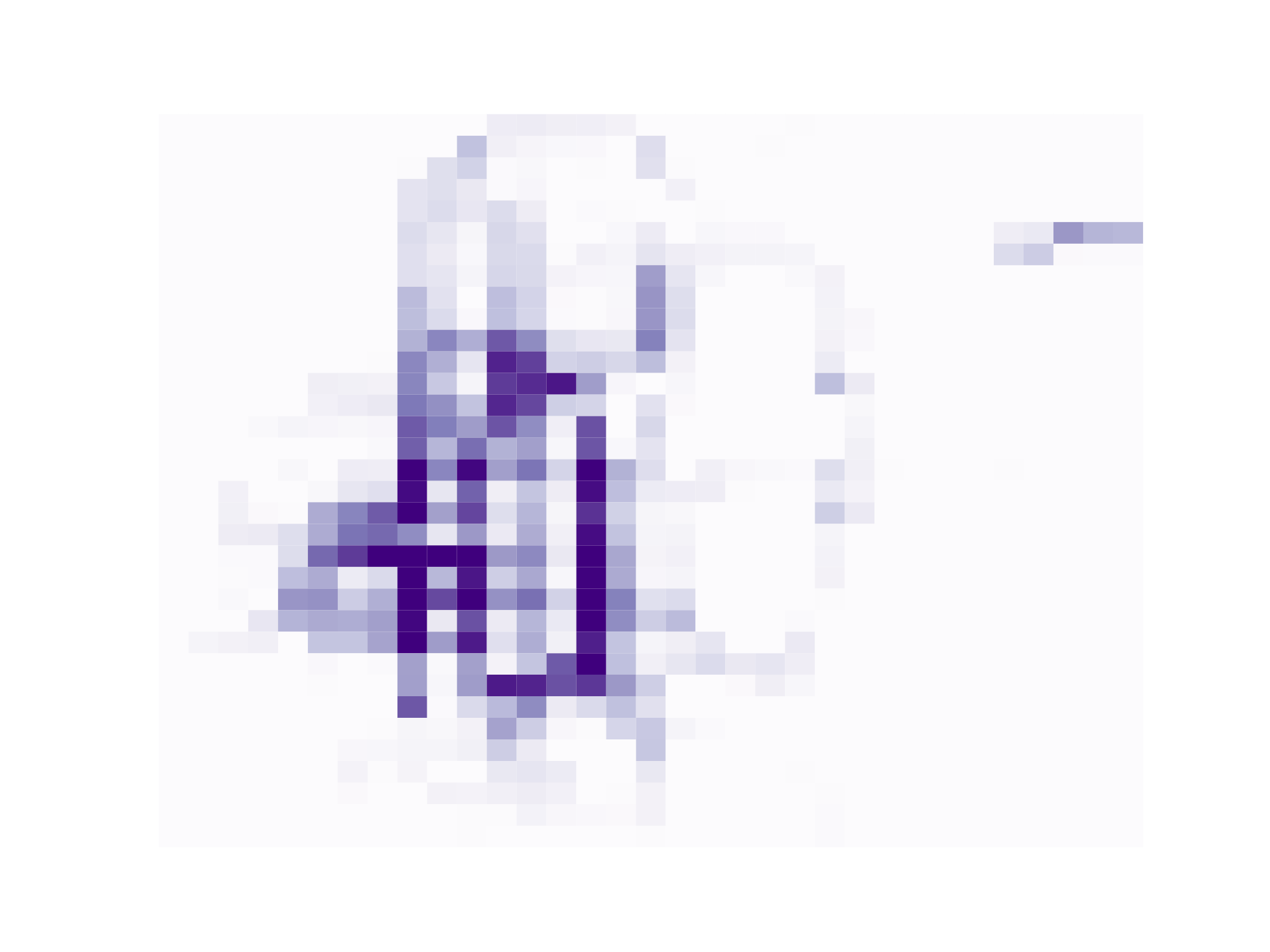}}
    \subfloat[\adatrace]{\includegraphics[width=0.24\textwidth,height = 0.14\textwidth]{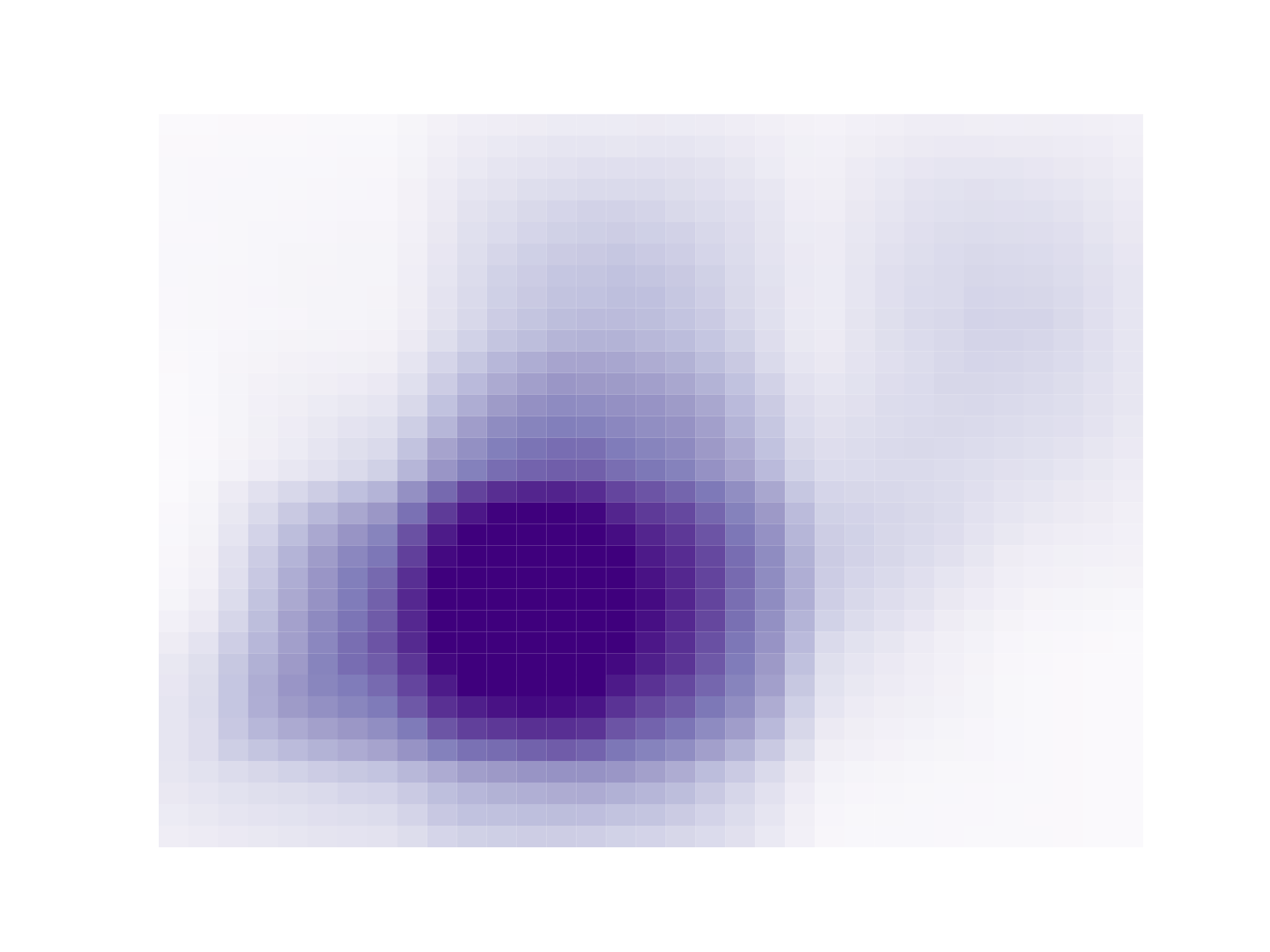}}
    \subfloat[\dpt]{\includegraphics[width=0.24\textwidth,height = 0.14\textwidth]{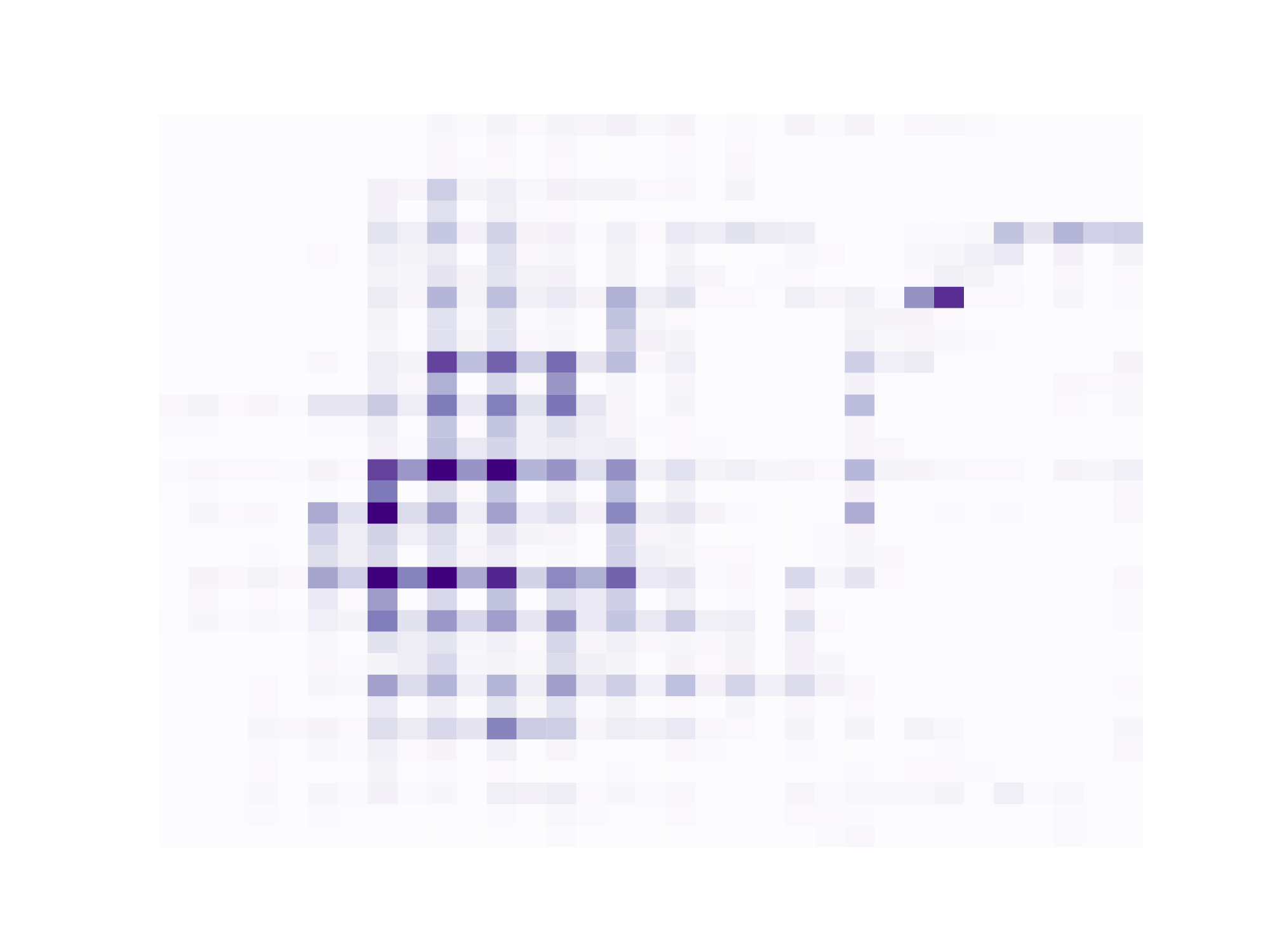}}
    \caption{Density visualization for different trajectory synthesis methods on the Taxi dataset.}

    \label{fig:visualization}
\end{figure*}

In this section, we provide an end-to-end evaluation to illustrate the effectiveness of \mymethod from two perspectives: quantitative evaluation and visualization.

\mypara{Quantitative Results}
We first quantitatively compare \mymethod with the state-of-the-art methods using the metrics discussed in \autoref{subsec:experimental_setup}.
\autoref{fig:comparison_private} illustrates the experimental results on three datasets. In general, we observe that \mymethod consistently outperforms \adatrace and \dpt for all settings.

For the length and diameter distribution, we observe that \mymethod reduces the \jsd by more than $50\%$ in most of the cases.
When the privacy budget is low, \mymethod can even achieve 1 order of magnitude performance improvement over the state-of-the-art.

With respect to the trajectory density and transition pattern, \mymethod can also reduce the error by at least $50\%$ in most settings.
We attribute this result to the benefit of the two-layer grids in the discretization process in \autoref{subsec:grid}.
Note that \adatrace also uses two-layer discretization.
As discussed in \autoref{subsec:grid}, the main difference is that \mymethod treat the second-layer cells as states while \adatrace only uses the second-layer grid for sampling.
The experimental results further validate that using the second-layer grid as states can capture more transition information, leading to better-quality synthetic trajectories.

\begin{table*}[!t]
    \footnotesize
    \caption{Details of ablation studies.
    In the first study (\autoref{fig:ablation1}), we evaluate component (a) and considers ablation 1-3; In the second study (\autoref{fig:ablation2}), we evaluate component (b) and considers ablation 3 and 4; In the last study (\autoref{fig:ablation3}), we evaluate component (c) and considers ablation 4 and 5.
    }
	\centering
	\begin{tabular}{c|c|c|c}
		\toprule
		Ablation study & Component (a) & Component (b) & Component (c) \\ 
		\midrule
		
		Ablation1 & First-order Markov model & -& -\\
		Ablation2 & Second-order Markov model  & -&-\\
		Ablation3 & Adaptive model & -&-\\
		Ablation4 & Adaptive model  &  Using second-layer cells as state&-\\
		Ablation5 & Adaptive model & Using second-layer cells as state & Trip distribution estimation\\
		
		\bottomrule
	\end{tabular}
	\label{tbl:ablation}
\end{table*}

\begin{figure*}[!htb]
    \centering
    
    \subfloat{\includegraphics[width=1\textwidth]{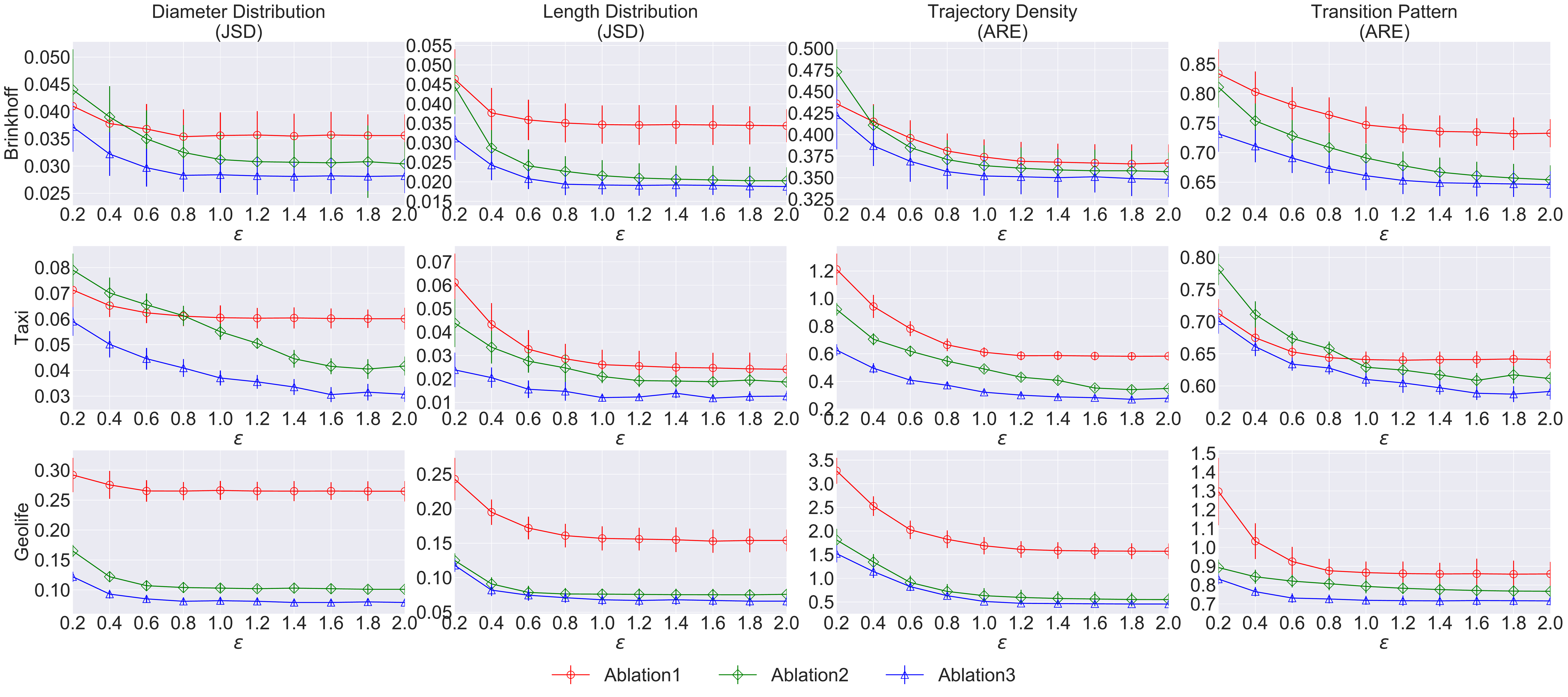}}
    \caption{Effectiveness of component (a).
    Ablation1 stands for the first-order model, Ablation2 stands for \mymethod with the second-order model, and Ablation3 stands for \mymethod with the adaptive model.
    All three methods are without components (b) and (c).}
    \label{fig:ablation1}
\end{figure*}

Interestingly, comparing \adatrace and \dpt on all datasets, we observe that \adatrace performs better on the Brinkhoff and Taxi datasets, while \dpt performs better on the Geolife dataset.
Furthermore, we observe that the absolute error values of \dpt are similar for all datasets, while \adatrace performs extremely poorly on Geolife.
After carefully checking the characteristics of all datasets, we find that the trajectory is concentrated in a small area on Geolife, while the trajectories in the other two datasets are more evenly distributed.
This observation indicates that \adatrace is not good at handling the datasets with trajectories concentrating in a small area on the map.
This can be explained by the fact that \adatrace only uses the first-layer cells as states, making most of the trajectories discretized into very short sequences of states.
Furthermore, since most of the trajectories are concentrated in a small number of cells, the transition patterns of trajectories lack diversity.
In this case, the \markovmodel and the trip distribution cannot learn useful information.

\mypara{Visualization Results}
To better illustrate the superiority of our proposed method \mymethod, we further provide a visualization comparison of the trajectory density in \autoref{fig:visualization}.
We experiment on the Taxi dataset and divide the map into an $80 \times 80$ uniform grid.
We use heatmaps to visualize the number of trajectories passing through each cell in the grid.
Darker cells means there are more trajectories passing through them.
The visualization results show that the trajectories generated by \mymethod have higher fidelity, while \adatrace and \dpt cannot preserve the density information.

\subsection{Ablation Study}
\label{exp:ablation_study}

\begin{figure*}[!htb]
    \centering
    
    \subfloat{\includegraphics[width=1\textwidth]{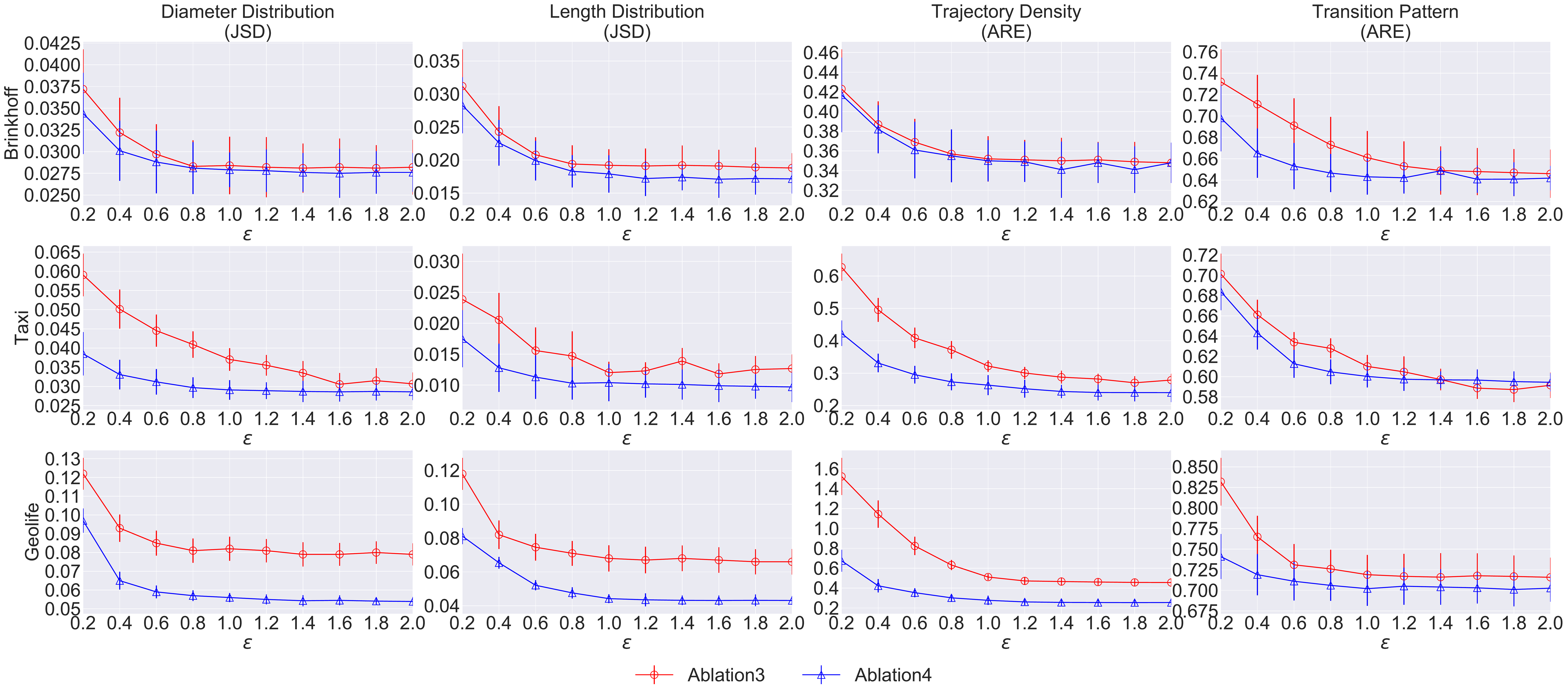}}
    \caption{Effectiveness of taking the second layer cells as states of \mymethod.
    Ablation3 stands for \mymethod without second layer cells, and Ablation4 stands for \mymethod with second layer cells.
    Both methods assume the adaptive model in component (a) but without component (c).
    }
    \label{fig:ablation2}
\end{figure*}

\begin{figure*}[!htb]
    \centering
    
    \subfloat{\includegraphics[width=1\textwidth]{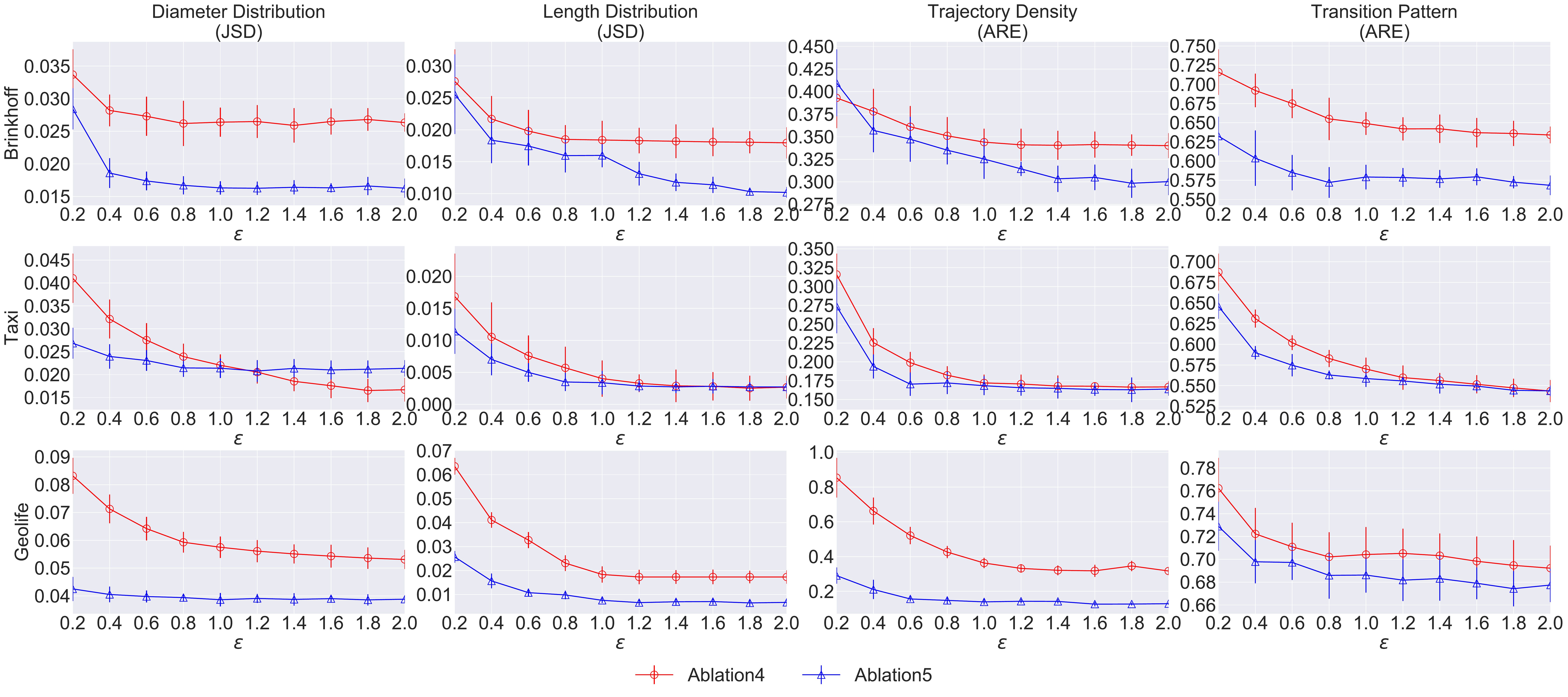}}
    \caption{Effectiveness of the trip distribution estimation of \mymethod.
    Ablation1 stands for \mymethod with adaptive model and (b), but without (c).
    Ablation5 stands for \mymethod with the adaptive model,  (b), and (c).
    }
    \label{fig:ablation3}
\end{figure*}

There are three main components in \mymethod: (a) the adaptive Markov models (b) taking the second-layer cells as states, and (c) the trip distribution estimation.
We conduct three ablation studies to evaluate the effectiveness of each of them in an incremental way, i.e., we first evaluate the effectiveness of (a), and then with the best option for (a), we evaluate (b), and then (c).

\mypara{Effectiveness of the Adaptive Markov Models}
First, we verify the effectiveness of the adaptive model with three variants called Ablation1, Ablation2, and Ablation3, which stand for three versions of \mymethod differing in component (a).
Specifically, Ablation1, Ablation2, and Ablation3 use the first-order Markov model, the second-order Markov model, and the adaptive model, respectively.
Meanwhile, all Ablation1, Ablation2, and Ablation3 are without (b) and (c).
The results in \autoref{fig:ablation1} show that Ablation3 outperforms the other two algorithms, which indicates that the adaptive model is the best choice among the three ways of using the Markov models.

\mypara{Effectiveness of Second-layer States}
Next, using the adaptive model in (a), we study the effectiveness of (b): How much do the second-layer cells help?
We have two methods, Ablation3 (without the second layer, from the last experiment) and Ablation4 (with the second layer), and the results are shown in \autoref{fig:ablation2}.
The results show that (b) is effective since Ablation4 outperforms Ablation3 for most datasets and metrics.

\mypara{Effectiveness of Trip Distribution}
The last evaluation is for component (c).
Similar to the last group, Ablation5 differs from Ablation4 by considering component (c) trip distribution estimation.
The results are shown in \autoref{fig:ablation3}.
From the results, we can conclude that (c) has a contribution to the accuracy since the performance of Ablation5 is better than Ablation4 in general.

\smallskip
For better reference, we include all the ablation variants in \autoref{tbl:ablation}.

\begin{figure}[!tb]
    \centering
    
    \subfloat{\includegraphics[width=0.5\textwidth]{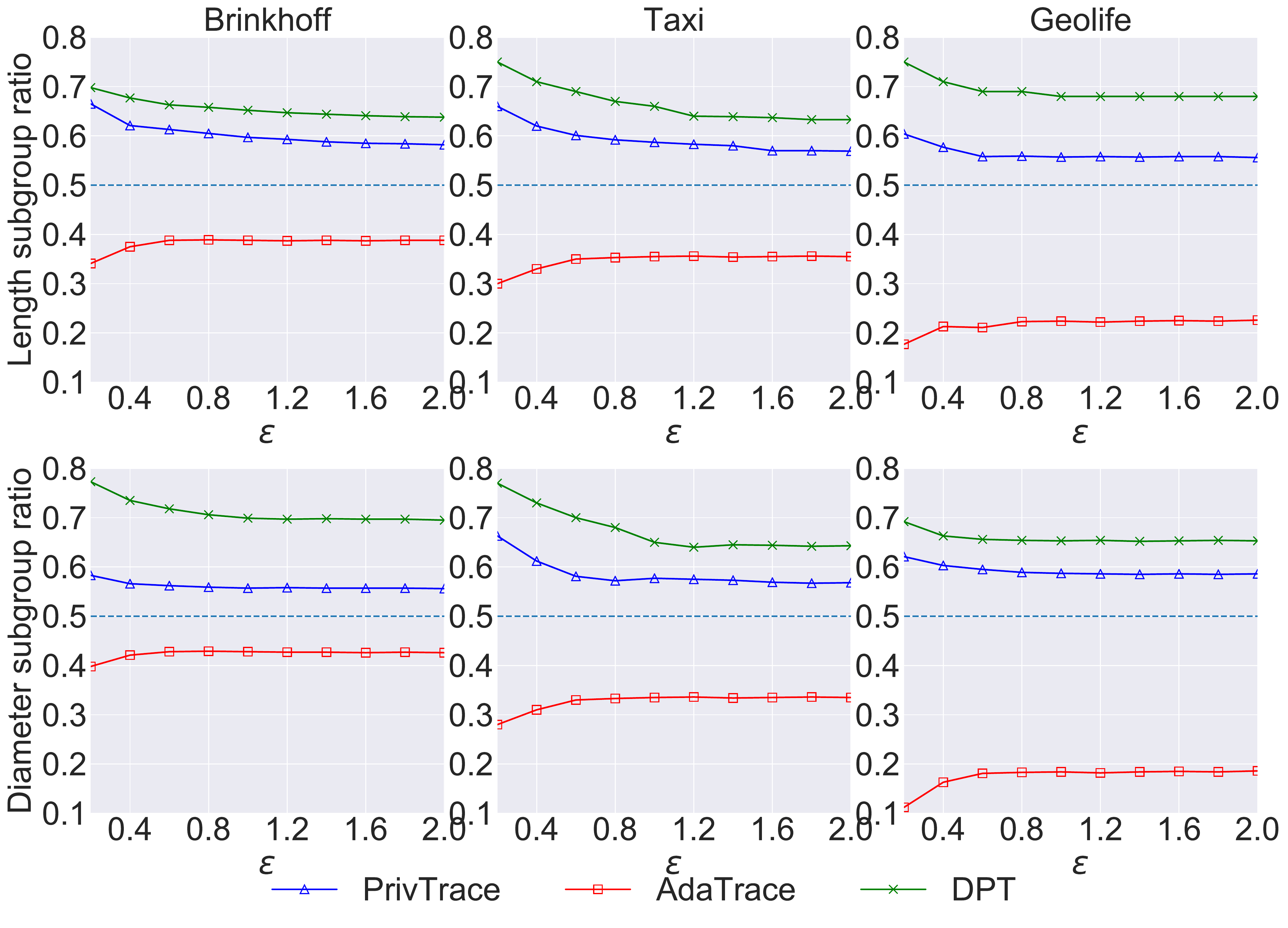}}

    \caption{Ratio of subgroup in different generation algorithms for different levels of $\epsilon$.
    The lines closer to 0.5 indicates less bias.}
    \label{fig:subgroup}
\end{figure}
\subsection{Impact on Subgroups}

Georgi et al.~\cite{ganev2021robin} focus on the impact of DP on the subgroups, and show that DP can affect the distribution and data utility of subgroups (more explanations about~\cite{ganev2021robin} can be found in
\autoref{subapp:subgroup}
).
Following the settings of~\cite{ganev2021robin}, we conduct two groups of evaluation to show the impact of DP trajectory data generation algorithms on subgroups.

\mypara{Subgroup Ratio}
First, we evaluate the impact of DP on the subgroup distribution for \mymethod, \adatrace, and \dpt.
In trajectory data, there is no data attribute or other demographic information such as gender that can be used in dividing subgroups directly.
Thus, we partition the trajectory data based on two common attributes: length and diameter.
We first calculate the median of the trajectory length in the original dataset, and then partition the dataset into the long subgroup $R_{l>l_{m}}$ and short subgroup $R_{l\leq l_{m}}$ based on the median value $l_{m}$ (thus 50\% trajectories are long and 50\% are short).
The diameter subgroups are defined similarly.

We compare the ratio of the short-length subgroup and the ratio of the short-diameter subgroup in the whole dataset for the synthetic data generated by \mymethod, \adatrace, and \dpt.
\autoref{fig:subgroup} illustrates the experimental results (each experiment is repeated 10 times).
In general, we observe that \mymethod has the least bias for the ratio of the subgroup.

\label{subsec:subgroups}

\begin{figure}[!tpb]
    \centering

    \subfloat{\includegraphics[width=0.5\textwidth]{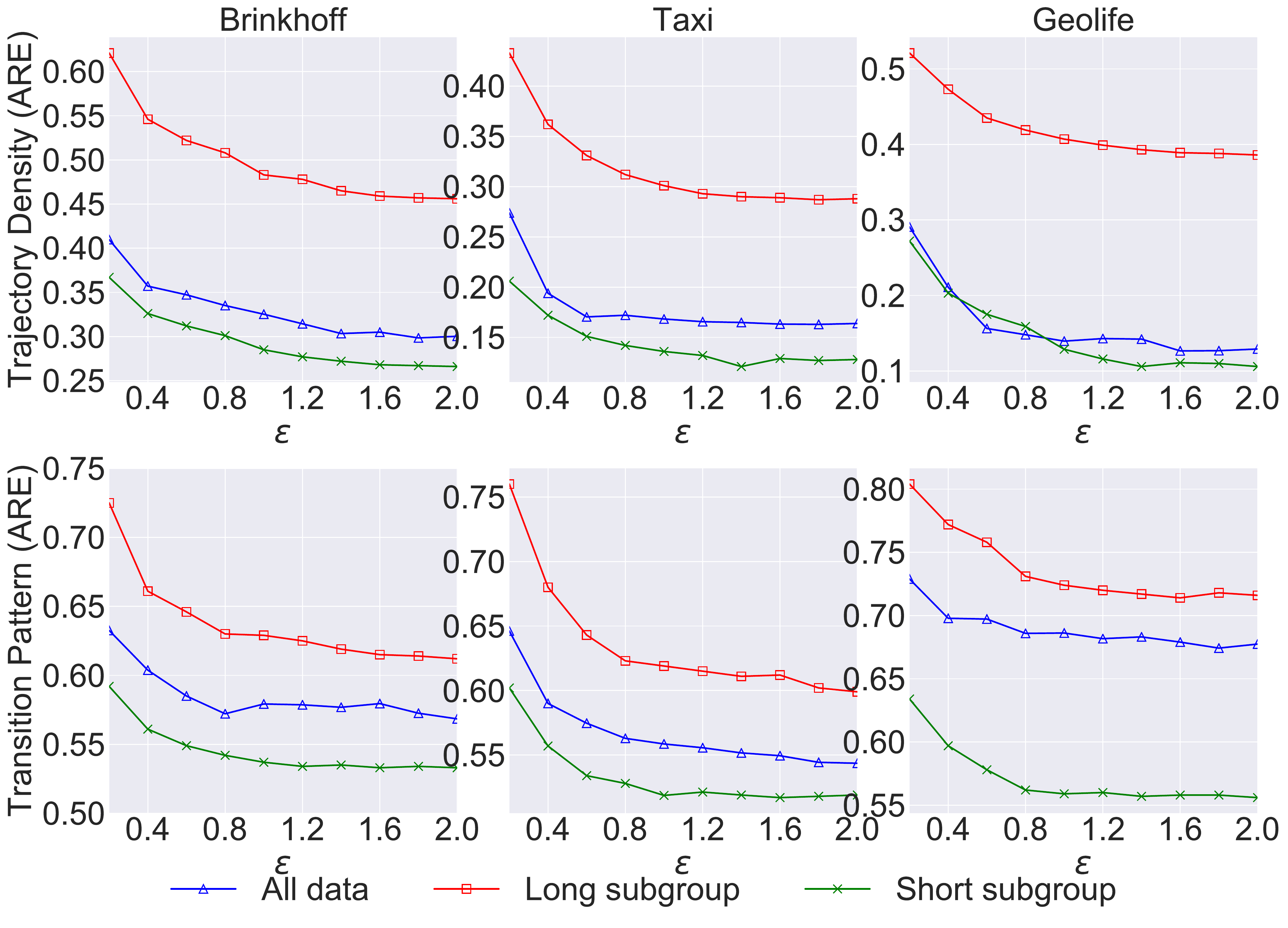}}
    \caption{Errors of the long subgroup and the short subgroup on the trajectory density and transition pattern metric.
    }
    \label{fig:ratio_utility_my}
\end{figure}

\mypara{Subgroup Utility}
Second, we evaluate the impact of \mymethod on the data utility of subgroups.
We use the median of length to divide subgroups and get four subgroups: The long subgroup from the original dataset $R_{l>l_{m}}$, the short subgroup from the original dataset $R_{l\leq l_{m}}$, the long subgroup from the synthetic dataset $S_{l>l_{m}}$, and the short subgroup from the original dataset $S_{l\leq l_{m}}$. 
We choose two metrics, i.e., the trajectory density metric and the transition pattern metric, to show the data utility.
In \autoref{subsec:end_to_end_evaluation}, these two metrics compare synthetic data with the original data.
However, it does not make sense if we compare the data in a subgroup with the whole original data directly.
Instead, we compare the subgroup from the original data and the synthetic data (such as $R_{l>l_{m}}$ and $S_{l>l_{m}}$), regarding the subgroup from the original dataset as the original data.
For a fair comparison, we sample the larger subgroup before comparison to make the two subgroups equal in size.

We conduct the experiments on two metrics and the conclusions for other metrics are consistent.
The experimental results in \autoref{fig:ratio_utility_my} show that the utility of the long subgroup is worse than that of the short subgroup.
We suspect this is due to the fact that long trajectories are more complicated and thus are more challenging to generate correctly.
The unbalanced utility between different subgroups is a limitation of \mymethod, as well as the limitation of \adatrace and \dpt (we defer the corresponding results to \autoref{subapp:subgroup} due to space limitation).

It is worth noting that our evaluations are only for the utility of subgroups, while the privacy-protection level is the same since all subgroups have the same privacy guarantee provided by DP.

%% file: 7_discussion.tex
\section{Discussion}
\label{sec:discussion}

\mypara{Time Consumption}
Theoretical analysis (\autoref{tbl:complexity} in \autoref{app:complexity_analysis}) shows that the time complexity of \mymethod is $\mathcal{O}(m  |\dset| + \statenumber^3)$, which is larger than \adatrace ($\mathcal{O}(m  |\dset| + m^2)$).
This is a limitation of \mymethod when $m$ is extremely large.
However, our empirical experiments show that the time consumption of \mymethod on real-world trajectory datasets is acceptable, e.g., less than 10 minutes (see \autoref{tbl:run_time} in \autoref{app:complexity_analysis} for more details).

\mypara{Practical Implementation}
With \mymethod (or other models), we can ensure the trajectories published/shared are protected by differential privacy. 
It is possible that given such a system, the data-miner/advertiser may make more predictions. This is not a limitation of the technique; instead, it is a consequence of users being attracted to share more data.

%% file: 8_related.tex
\section{Related Work}
\label{sec:relatedwork}

\mypara{Location Density Estimation}
There are a number of previous studies focusing on estimating the density distribution of a location dataset while satisfying DP.
Most approaches rely on recursive partitioning, which recursively performs binary partitioning on the map.
Xiao et al.~\cite{xiao2010differentially} propose to use KD-tree, which recursively partitions the map along some dimensions. 
In order to minimize the non-uniformity error, the authors use the heuristic to choose the partition point such that the two sub-regions are as close to uniform as possible.
To improve the estimation accuracy, Cormode et al.~\cite{cormode2012differentially} propose several alternatives based on KD-trees.
Instead of using a uniformity heuristic, they partition the nodes along the median of the partition dimension. 
The height of the tree is predetermined, and the privacy budget is divided among the levels.
The key challenge of the recursive partitioning-based approaches lies in choosing the right partition granularity to balance the noise error and the non-uniformity error. 
To address this issue, Qardaji et al.~\cite{dpgrid2013} propose a uniform-grid approach, which applies an equal-width grid of a certain size over the data domain and then issues independent count queries on the grid cells.

\mypara{Trajectory Data Publication}
Several studies have been done on finishing certain tasks using trajectory data or publishing trajectory data via perturbing locations.
Chen et al. \cite{chen2012differentially} propose to use differentially private prefix tree to publish transition data.
They established a method to generate transition data using a differentially private prefix tree.
They evaluated its mechanism by studying the impact of its protection on range query and sequential pattern mining.
Acs et al. \cite{acs2014case} propose a mechanism to publish spatio-temporal trajectory density data, which is counts of active users within small areas for given time windows.
Jiang et al. \cite{jiang2013publishing} propose to protect ships’ trajectories by location perturbing.
They preserve the endpoints of trajectories while the intermediate locations are altered by adding some noise satisfying differential privacy guarantees.
Also, there are some recent works focusing on generating synthetic trajectory data following other privacy requirements instead of different privacy. One of the most famous works is SGLT \cite{bindschaedler2016synthesizing}. SGLT uses plausible deniability as the privacy requirement. It captures both geographic and semantic features of real traces and uses the real traces are seeds to generate synthetic trajectories.

Our paper focuses on a more general paradigm that publishes a synthetic trajectory dataset while satisfying DP.

\mypara{Differentially Private Tabular Data Synthesis}
For the general tabular data, there exists a number of studies following the same paradigm of our paper, i.e., generating tabular data while satisfying DP.
They can be broadly classified into three categories: graphical model-based methods, game-based methods, and deep generative model-based methods.

The graphical model-based methods aim to estimate a graphical model that approximates the distribution of the original dataset in a differentially private manner and generate synthetic data by sampling from it.
For instance, PrivBayes~\cite{zhang2017privbayes} and BSG~\cite{bindschaedler2017plausible} adopt Bayesian network to approximate the data distribution, while PGM~\cite{mckenna2019graphical} and JTree~\cite{chen2015differentially} use Markov random field to approximate the data distribution.
PrivSyn~\cite{zhang2022privsyn} utilizes a number of marginal tables to capture as much as correlated information in the dataset, which can be regarded as a dense graphical model.
The core idea of game-based methods is to formulate the dataset synthesis problem as a zero-sum game~\cite{hardt2012simple,gaboardi2014dual,vietri2020new}.
Assume there are two players, a data player and a query player.
MWEM~\cite{hardt2012simple} method solves the game by having the data player use a no-regret learning algorithm, and the query player repeatedly best responds.
Dual Query~\cite{gaboardi2014dual} switches the role of the two players.
The deep generative model-based approaches rely on the DP-SGD framework~\cite{abadi2016deep} (adding noise in the optimization procedure) to train a generative model, and use this deep generative model to generate a synthetic dataset.
The most commonly used deep generative model is the Generative Adversarial Network (GAN)~\cite{zhang2018differentially,beaulieu2019privacy,abay2018privacy,frigerio2019differentially,tantipongpipat2019differentially}.

%% file: 9_conclusion.tex
\section{Conclusion}
\label{sec:conclusion}

In this paper, we propose a differentially private algorithm to generate trajectory data.
By employing the first-order and second-order \markovmodels, we achieve a middle ground between richness of information and amount of noise.
Besides, we propose an optimization-based method to estimate the trip distribution, which is important information for generating synthetic trajectory data.
Extensive experiments on real-world and generated datasets are conducted to illustrate the superiority over the current state of the art.

\section*{Acknowledgments} 
We thank the anonymous shepherd and reviewers for their constructive feedback.
This work is supported by ``New Generation Artificial Intelligence'' major project of China under Grant 2018AAA0101605, National Natural Science Foundation of China (NSFC) under grant NO.61833015, NSF CNS-2220433, CNS-2213700, CCF-2217071, and the Helmholtz Association within the project ``Trustworthy Federated Data Analytics'' (TFDA) (funding number ZT-I-OO1 4).

%% file: algorithm_analysis.tex
\section{Computational Complexity Analysis}
\label{app:complexity_analysis}

In this section, we theoretically analyze the computational complexity of different methods and then empirically evaluate their running time and memory consumption.

Here we assume a method discretizes the space into $m_i$ cells in the $i$-th layer and the total number of cells $m=\sum m_i$.  We also assume that the average length of the trajectory is a constant.

\mypara{Time Complexity}
The time complexity of \mymethod can be analyzed by studying every step of the algorithm.
In the discretization step, \mymethod discretizes the geographical space into $\statenumber$ cells and translates trajectories from location sequences to state sequences (i.e., find the cells where a trajectory has passed for all the trajectories).
To find if a trajectory passes a certain cell takes $\mathcal{O}(\statenumber  |\dset|)$.
The process of calculating a Markov chain model requires counting the transitions in every trajectory.
Thus, the time complexity of calculating Markov chain models is $\mathcal{O}(|\dset|)$.
The trip distribution estimation is composed of calculating the shortest path and solving the convex optimization problem.
The time complexities of them are $\mathcal{O}(\statenumber^2\ln(\statenumber))$ using the Dijkstra's algorithm and $\mathcal{O}(\statenumber^2)$, respectively.
The time complexity of generating and adding noise for the first-order and second-order models are $\mathcal{O}(\statenumber^2)$ and $\mathcal{O}(\statenumber^3)$.
Finally, generating a trajectory depends on the transition count distributions in the Markov chain models, which has time complexity is $\mathcal{O}(|\dset|)$.
Putting all steps together, the time complexity of \mymethod is $\mathcal{O}(\statenumber  |\dset|) + \mathcal{O}(|\dset| + \statenumber^2\ln(\statenumber)) + \mathcal{O}(|\dset|) + \mathcal{O}(\statenumber^2) + \mathcal{O}(\statenumber^3) = \mathcal{O}(\statenumber  |\dset| + \statenumber^3)$.

The analysis for \adatrace is similar to \mymethod: 
The time complexity of discretization is $  \mathcal{O}(m|\dset|)$.
The time complexity for learning the Markov chain model and estimating trip distribution are both $\mathcal{O}(|\dset|)$ since it involves scanning all trajectories.
In the length distribution calculating, \adatrace scans all trajectories and then chooses a proper distribution for all possible trips.
The time complexity of length distribution calculating is $\mathcal{O}(|\dset| + m^2)$.
In the steps above, the Laplacian noise is added to cell density, the first-order Markov model, and trip counts, respectively.
The time complexity of Laplacian noise adding in \adatrace is $\mathcal{O}(m) + \mathcal{O}(m^2) + \mathcal{O}(m^2) = \mathcal{O}(m^2)$.
As the last step, the random walk process takes $\mathcal{O}(m|\dset|)$ to generate synthetic trajectories.
The total time complexity of \adatrace is $\mathcal{O}(m  |\dset| + m^2)$.

For \dpt, the discretization step uses several grids with different granularities (which are called hierarchical reference systems) to discretize the geographical space.
In \dpt, when the granularity of the division becomes finer, the number of states increases exponentially.
We denote $g$ as the number of different granularities and $m_g$ as the number of states in the most fine-grained grid.
The time complexity of discretization step is $\mathcal{O}(m_g|\dset|    )$.
In the prefix tree-building step, the algorithm takes $\mathcal{O}(|\dset|)$ for scanning all trajectories.
After the prefix trees-building, the Laplacian noise will be injected to every node of the trees.
The noise adding takes time complexity as $\mathcal{O}(m_g^{h})$, where $h$ is the height of the tree with the most fine-grained grid. 
In the data generation process, \dpt employs random walk on prefix tree and takes $\mathcal{O}(|\dset|)$.
The total time complexity is $\mathcal{O}(m_g|\dset|) + \mathcal{O}(m_g^{h}) = \mathcal{O}(m_g|\dset| + m_g^{h})$.

\begin{table}[!t]
    \caption{Comparison of computational complexity for different methods.}
	\centering
	\begin{tabular}{c|c|c}
		\toprule
		Methods & Time Complexity & Space Complexity \\ 
		\midrule
		
		\adatrace~\cite{gursoy2018utility} & $\mathcal{O}(m |\dset| + m^2)$ & $\mathcal{O}(m_1^2)$ \\
		\dpt~\cite{he2015dpt} & $\mathcal{O}(m_g|\dset| + m_g^{h})$ & $\mathcal{O}(m_g^{h})$ \\
		\mymethod & $\mathcal{O}(m |\dset| + \statenumber^3)$ & $\mathcal{O}(\statenumber^3)$ \\
		\bottomrule
	\end{tabular}
	\label{tbl:complexity}
\end{table}

\begin{table}[!t]
    \caption{Comparison of running time for different methods.}
	\centering
	\begin{tabular}{c|c|c|c}
		\toprule
		Datasets & Brinkhoff & Taxi & Geolife \\ 
		\midrule
		
		\adatrace~\cite{gursoy2018utility} & 2 min 30 s & 4 min 44 s & 1 min 45 s\\
		\dpt~\cite{he2015dpt} & 18 min 47 s & 43 min 51 s & 34 min 27 s \\
		\mymethod & 6 min 21 s & 8 min 33 s &  7 min 24 s\\
		\bottomrule
	\end{tabular}
	\label{tbl:run_time}
\end{table}

\begin{table}[!t]
    \caption{Comparison of memory consumption of different methods.
    The unit is Megabytes.}
	\centering
	\begin{tabular}{c|c|c|c}
		\toprule
		Datasets & Brinkhoff & Taxi & Geolife \\ 
		\midrule
		
		\adatrace~\cite{gursoy2018utility} & 0.05 & 0.04 & 0.53\\
		\dpt~\cite{he2015dpt} & 1003.77 & 111.22 & 1755.23 \\
		\mymethod & 12.96 & 63.30 & 193.75 \\
		\bottomrule
	\end{tabular}
	\label{tbl:space_consumption}
\end{table}

\mypara{Space Complexity}
The memory consumption of all methods mainly comes from the model building process.
\mymethod uses the first-order Markov chain model and the second-order Markov chain model in model building.
The space complexity of \mymethod is $\mathcal{O}(\statenumber^3)$.

\adatrace stores the first-order Markov chain model, trip distribution, and length distribution in memory.
For the first-order Markov chain model, there are $m_1^2$ transition count values to be stored.
For trip distribution and length distribution, information for all possible trips is stored and takes $\mathcal{O}(m_1^2)$ space.
The space complexity for \adatrace is $\mathcal{O}(m_1^2)$.

\dpt stores prefix trees in memory.
The number of nodes in the tree of the most fine-grained grid is $m_g  $.
The memory consumption of this tree is $\mathcal{O}(m_g^{h})$, where $h$ is the height of this tree.
Since the most fine-grained tree consumes the most memory, the space complexity of \dpt is $\mathcal{O}(m_g^{h})$.
Both the time and space complexity of different algorithms are shown in \autoref{tbl:complexity}.

\mypara{Empirical Evaluation}
\autoref{tbl:run_time} and \autoref{tbl:space_consumption} illustrate the running time and memory consumption for all methods on the three datasets.
The empirical running time in \autoref{tbl:run_time} shows that \adatrace has the best running time performance. 
This is because \adatrace has relatively low time complexity as shown in \autoref{tbl:complexity}.
\mymethod is slower than \adatrace since it contains a graph-based method and convex optimization in the trip distribution estimation.
For memory consumption in \autoref{tbl:space_consumption}, we observe that \dpt consumes the most memory since storing the reference system is space-consuming.

%% file: app_method.tex
\section{More Details}

\subsection{Details of Normalization in \autoref{subsec:grid}}
\label{app:method_grid}

Denote the number of occurrences of trajectory $\realtrajectory_{i}$ in cell $j$ as $\delta_{ij}$, we can represent the trajectory density as a vector $(\sum_{i} \delta_{i1}, \sum_{i} \delta_{i2}, \dots \sum_{i} \delta_{iK^{2}})$.
The sensitivity of the trajectory density is the impact of one trajectory on this vector, which is unbounded since one trajectory could appear in any number of cells.
To bound the sensitivity, we propose to use the length-normalized density, which is defined by $\delta^{'}_{ij} = \delta_{ij} / \sum_{j} \delta_{ij}$ instead of $\delta_{ij}$.
Denote the length-normalized density query vector $(\sum_{i} \delta^{'}_{i1}, \sum_{i} \delta^{'}_{i2}, \dots \sum_{i} \delta^{'}_{iK^{2}})$ as $\densityvec(\dset)$, we can see

\begin{equation*}
    \begin{aligned}
        \mathsf{GS}_{\densityvec} &=
        \max\limits_{\realtrajectory_{k} \in \dset} ||\densityvec(\dset) - \densityvec(\dset - \realtrajectory_{k})||_1\\
        &=\max\limits_{\realtrajectory_{k} \in \dset}||(\delta^{'}_{k1}, \delta^{'}_{k2}, \dots, \delta^{'}_{k\levelonedipar^2})||_1\\
        &= \max\limits_{\forall \realtrajectory_{k} \in \dset}|\sum\limits_{j=1}^{\levelonedipar^2} \delta_{kj} / \sum\limits_{j=1}^{N^2} \delta_{kj}|
        = 1
    \end{aligned}
\end{equation*}

\subsection{Details of Bounding Sensitivity in \autoref{subsec:markov_model_building}}

\label{pf:transition_model_sensitivity}

In \markovmodel, denote $\sigma_{i}$ as a state corresponding to $\cellinmodel_{i}$, and any trajectory is transformed into a state sequence.
For example, if a trajectory $\realtrajectory$ passes through the following cells in order: $\cellinmodel_{2}, \cellinmodel_{3}, \cellinmodel_{7}$, the state sequence it is transformed into is $(\markovstate_{2}, \markovstate_{3}, \markovstate_{7})$.

We denote the length-normalized count for transition $\transition$ as $N_{\dset}^{'}(\transition) = \sum_{\forall \stateseq \in \dset} \frac{N_{\stateseq}(\transition)}{|\stateseq|}$, where $N_{\stateseq}(\transition)$ means the appearance of $\transition$ in trajectory $\stateseq$.
And the set of length-normalized transition counts is defined as $\marquery^{\marorder}(\dset) = \{N_{\dset}^{'}(\transition), \forall \transition \in \Sigma^{\marorder+1}\}$.
Applying the length-normalization technique, the $\marorder$th-order \markovmodel can be built easily by calculating $\marquery^{\marorder}$.

Here the core idea of bounding sensitivity is to divide the transition count by the length of the trajectory.
By doing this, the sensitivity of the \markovmodel counting query can be bounded by 1.

\begin{theorem}
\label{thm:transition_model_sensitivity}
Define $N_{\dset}^{'}(r) = \sum_{\forall \stateseq \in \dset} \frac{N_{\stateseq}(r)}{|\stateseq|}$ where $r$ is any length-$\marorder+1$ sequence, the sensitivity of outputing all $N_{\dset}^{'}(r)$, denoted by $\marquery^{\marorder}(\dset) = \{N_{\dset}^{'}(r), \forall r \in \Sigma^{\marorder+1}\}$, is 1.
\end{theorem}

The theorem can be proved by first showing that a single sequence $\stateseq$'s contribution $\marquery^{\marorder}$ is limited by $|\stateseq|$.
The following lemma tells us that $\sum_{r \in \Sigma^{l}} N_{\stateseq}(r)$ cannot be larger than $|\stateseq|$.
\begin{lemma}
Given a set of all possible subsequences of length $l$, denoted by $\Sigma^{l}$, and a sequence $\stateseq$, $\sum_{r \in \Sigma^{l}} N_{\stateseq}(r) \leq |\stateseq|$, where $|\stateseq|$ is the length of sequence $\stateseq$.
\label{lemma:length_div}
\end{lemma}

\label{pf:length_div}
\begin{proof}

The calculating of $\sum\limits_{r \in \Sigma^{l}} N_{\stateseq}(r)$ is to count all subsequences of length $l$ in $\stateseq$. Considering the sequence can only have no more than $|\stateseq|$ continuous subsequences of the same length, $\sum\limits_{r \in \Sigma^{l}} N_{\stateseq}(r) \leq |\stateseq|$.
\end{proof}

With that, we can prove \autoref{thm:transition_model_sensitivity}:

\begin{proof}
We denote a dateset of sequences as $\dset$.
The sensitivity of $\marquery^{\marorder}$ is $\max_{\forall \stateseq^{'} \in \dset} \marquery^{\marorder}(\dset) - \marquery^{\marorder}(\dset - \stateseq^{'})$.
We give an order to all transitions $\transition$ of length $\marorder + 1$ and $\marquery^{\marorder}$ is represented as a vector following the order of transitions.
For a transition $\transition_{i}$, the ith element of $\marquery^{\marorder}(\dset) - \marquery^{\marorder}(\dset - \stateseq^{'})$ is $N_{\dset}^{'}(\transition_{i}) - N_{\dset - \stateseq^{'}}^{'}(\transition_{i}) = N_{\stateseq^{'}}^{'}(\transition_{i})$.
Thus, the sensitivity is,
\begin{equation*}
    \begin{aligned}
        \mathsf{GS}_{\marquery^{\marorder}} &=\marquery^{\marorder}(\dset) - \marquery^{\marorder}(\dset - \stateseq^{'})\\
        &=\max\limits_{\forall \stateseq^{'} \in \dset}||(N_{\stateseq^{'}}^{'}(\transition_{1})), \dots, (N_{\stateseq^{'}}^{'}(\transition_{\statenumber^{\marorder + 1}}))||_1\\
        & = \max\limits_{\forall \stateseq^{'} \in \dset}|\sum\limits_{\transition_{i}}N^{'}_{\stateseq^{'}}(r)|\\
          & = \max\limits_{\forall \stateseq^{'} \in \dset}\sum\limits_{\transition_{i}}\frac{N_{\stateseq^{'}}(r)}{|\stateseq^{'}|}
    \end{aligned}
\end{equation*}

According to \lemmaref{lemma:length_div}, $\sum\limits_{r \in \Sigma^{l}} N_{r}(\stateseq) \leq |\stateseq|$, $\mathsf{GS}_{N^{'}} \leq 1$
\end{proof}

\subsection{Details of Trip Distribution Estimation in \autoref{subsec:optimization}}
\label{subapp:method_optimization}

\mypara{Calculation for $\countstart_{i}$}
Considering the sets of states $\markovstate_{1}, \dots, \markovstate_{\statenumber}$, by calculating $\marquery^{1}$ we will get $\transitioncount^{'}_{\dset}((\markovstate_{start},\markovstate_{i}))$ and $\transitioncount^{'}_{\dset}((\markovstate_{j},\markovstate_{end}))$ for all $\markovstate_i$ and $\markovstate_j$.
For presentation purpose, we denote $\transitioncount^{'}_{\dset}((\markovstate_{start},\seqstate_{i}))$ and $\transitioncount^{'}_{\dset}((\seqstate_{j},\markovstate_{end}))$ as $\countstart^{'}_{i}$ and $\countend^{'}_{j}$, respectively.
$\countstart^{'}_{i}$ and $\countend^{'}_{j}$ are values of $\countstart_{i}$ and $\countend_{j}$ before adding Laplacian noise.
Relation between $\tripset_{ij}$ and $\countstart_{i}$ can be built through $\countstart^{'}_{i}$.

Without loss of generality, we first analyze the calculation of $\countstart^{'}_{i}$ ($\countend^{'}_{j}$ can be calculated in a similar way).
We denote the set of all trajectories starting at $\markovstate_{i}$ and ending at $\markovstate_{j}$ by $\tripset_{ij}$. 
$\frac{|\tripset_{ij}|}{\sum_{\statetrajectory \in \tripset_{ij}}1/|\statetrajectory|}$ is the harmony average of the sequence length of trajectory in $\tripset_{ij}$.
With the assumption that people tend to travel the shortest path, we have $\harlength_{ij}$ is very close to the harmony average length.
In later calculation, we use $\harlength_{ij}$ as the harmony average length.
Here we provide how we compute the relation $\countstart^{'}_{i} = \sum_{j} \frac{\tripnum_{ij}}{\harlength_{ij}}$:
$\countstart^{'}_{i}$ is the length-normalized transition count of transition $\transition_{i}^{s}=(\markovstate_{start},\markovstate_{i})$, which is calculated by $\countstart^{'}_{i} = N_{\dset}^{'}(\transition_{i}^{s})
        =\sum_{\forall \statetrajectory \in \dset} \frac{N_{\statetrajectory}(\transition_{i}^{s})}{|\statetrajectory|}=\sum_{j} \sum_{i} \sum_{\forall \statetrajectory \in \tripset_{ij}} \frac{N_{\statetrajectory}(\transition_{i}^{s})}{|\statetrajectory|}$.
Considering that every trajectory can only belongs to one $\tripset$, we can divide $\dset$ into two sets:$\tripset_{ij}$ and $\dset-\tripset_{ij}$.
By this division, we have
        $\countstart^{'}_{i} =\sum_{j} \sum_{\forall \statetrajectory \in \tripset_{ij}} \frac{N_{\statetrajectory}(\transition_{i}^{s})}{|\statetrajectory|}+ \sum_{j} \sum_{v \neq i} \sum_{\forall \statetrajectory \in \tripset_{vj}} \frac{N_{\statetrajectory}(\transition_{i}^{s})}{|\statetrajectory|}$.
It can be concluded that $N_{\statetrajectory}(\transition_{i}^{s})=1$ if $\statetrajectory$ starts at $\markovstate_{i}$ and $N_{\statetrajectory}(\transition_{i}^{s})=0$ otherwise, since a trajectory only has one start state.
Thus, for $\statetrajectory$ in $\tripset_{ij}$, $N_{\statetrajectory}(\transition_{i}^{s})=1$.
By the definition of the harmony average and replacing $|\tripset_{ij}|$ by $\tripnum_{ij}$ , we have $\sum_{\forall \statetrajectory \in \tripset_{ij}} \frac{1}{|\statetrajectory|} = \frac{\tripnum_{ij}}{\harlength_{ij}}$.
Replacing $\sum_{\forall \statetrajectory \in \tripset_{ij}} \frac{1}{|\statetrajectory|}$ by $\frac{\tripnum_{ij}}{\harlength_{ij}}$,
 we have $\countstart^{'}_{i} = \sum_{j} \sum_{\forall \statetrajectory \in \tripset_{ij}} \frac{1}{|\statetrajectory|} + 0 = \sum_{j} \frac{\tripnum_{ij}}{\harlength_{ij}}$.
Since $\countstart$ is $\countstart^{'}$ together with the Laplacian noise, we have $\countstart_i\simeq \sum\limits_{j=1}^m \frac{t_{ij}}{\harlength_{ij}}$.

\mypara{Estimating $\harlength_{ij}$}
Observing that the shortest length trajectory is similar to the minimum-weight shortest path in a graph, we design a graph-based method to estimate $\harlength_{ij}$.
We construct a graph whose nodes are states in the \markovmodel.
An edge exists between two nodes if the two cells corresponding to the two states are neighbors.
The weight of an edge is the geographical distance between the two cells.
Then, we use the minimum-weight shortest path algorithm to calculate $\harlength_{ij}$.

\subsection{Details of Evaluation Metrics in \autoref{subsec:experimental_setup}}
\label{subapp:evaluation_metrics}

\begin{itemize}[leftmargin=*]
\setlength\itemsep{-0.25em}
\vspace{-0.5mm}
    \item \mypara{Length Distribution}
    The length of a trajectory measures the summation of distances between all two adjacent points.
    We bucketize the length into 50 bins and count the number of trajectories falling into each bin to calculate the length distribution.
    We use the Jensen-Shannon divergence (\jsd) to measure the error between $\dset_s$ and $\dset_o$.
    
    \item \mypara{Diameter Distribution}
    The diameter indicates the maximum distance between any two points in a trajectory.
    Similar to the length distribution, we bucketize the diameter into 50 bins to obtain the diameter distribution and use \jsd to measure the error.

    \item \mypara{Trajectory Density}
    It measures the number of trajectories passing through a specific area.
    We first generate 500 random circle areas with random radius in the map, and count the number of trajectories passing through each area.
    We use the average relative error (\are) to measure the error.
    Denote the set of density queries as $Q$,
    the \are of all queries is calculated as:
    \vspace{-1mm}
    \begin{equation*}
        \are_{TD}=\frac{1}{|Q|} \sum_i \frac{|Q_i(\dset_o) - Q_i(\dset_s)|}{max(Q_i(\dset_o), \smallvalueinexp)}
    \end{equation*}
    where $|Q|$ is the cardinality of $Q$, $\smallvalueinexp$ is a factor to bound the impact of a query of small real value. 

\item \mypara{Transition Pattern}
    It captures the frequency of transiting from one place to another.
    We use a $20 \times 20$ uniform grid to discretize the geographical space and count the frequencies of all transition patterns.
    In most downstream tasks, only frequent patterns are considered.
    We use the \are of the top $\mu$ frequent patterns to measure the error.
    The \are of the top-$\mu$ frequent transition patterns $\transition$ is calculated as:
    \begin{equation*}
        \are_{TP}=\frac{1}{\mu} \sum_{P \in \mathcal{P}_{top}} \frac{|\transitioncount_{\transition}(\dset_o) - \transitioncount_{\transition}(\dset_s)|}{max(\transitioncount_P(\dset_o), \smallvalueinexp)}
    \end{equation*}
    Here $\transitioncount_P$ means the number of pattern $P$’s occurrences and $\mathcal{P}_{top}$ means the set of the top-$\mu$ frequent transition patterns.
    In our experiments, we only consider the transition patterns with length between 2 to 5, i.e., the transition patterns whose number of states is between 2 to 5.
    Then we rank these counts and find the top-$\mu$ transition patterns since users usually focus on the frequent patterns.
    In the experiments, we set $\mu$ as 200.
\end{itemize}

\subsection{Details of Competitors in \autoref{subsec:experimental_setup}}
\label{appsub:competitors}

The \adatrace code\footnote{\url{https://github.com/git-disl/AdaTrace}} chooses between three candidate distributions by comparing the candidate distributions with distribution in the original dataset without adding noise (in LengthDistribution.java, Line 133 to Line 135).
Meanwhile, \adatrace extract mean and median from original dataset using the same privacy budget (in LengthDistribution.java, line 81 and 86).
For a fair comparison, we fix this problem and report the results of the fixed version of \adatrace.
That is, we divide the privacy budget for length distribution extraction in \adatrace into three parts equally.
The process of extracting mean and median and choosing a distribution from the candidate consumes one part, respectively.

\subsection{Details of NormCut in Section 4.3}
\label{subapp:negative_value}

We adopt the postprocessing method NormCut~\cite{wang2020locally,zhang2022privsyn} to handle the negative values.
The general idea of NormCut is to make the negative values zero, tally the amount (sum of the abstract values of the negative values), and renormalize (subtract an equal amount from the positive values).

\begin{algorithm}[!t]
\SetAlgoLined
\LinesNumbered
\KwIn{
A vector $v$ with negative value}
\KwOut{A non-negative vector $v^{'}$}

Calculate the sum of all negative values in $v$ as $N$;

Calculate the sum of all positive values in $v$ as $P$;

\While{$N \neq 0$ and $P \neq 0$}{
\label{In2:rw1}

    Find the smallest positive value in $v$ as $p_{s}$.
    
    $temp = p_{s} + N$.
    
    \eIf{$temp < 0$}{
    
    $N = temp$
    
    $P = P - p_{s}$
    
    $p_{s} = 0$
    }{
    $N = 0$
    
    $p_{s} = temp$
    \label{In2:rw2}
    }
}

Turn all existing negative values in $v$ to 0.
\caption{NormCut}
\label{alg:norm-cut}
\end{algorithm}

\section{Detailed Proof of \mymethod's DP Guarantee}
\label{app:dp_proof_detials}

\begin{proof}
\mymethod consists of four components: Geographical space discretization, Markov chain models learning, trip distribution estimation, and trajectories generation.
In the following, we check the privacy budget consumption of each component.

\begin{itemize}[leftmargin=*]
\item \textbf{Geographical Space Discretization.}
    We first count the occurrences of trajectories passing through the cells in the coarse-grained grid. We then add Laplacian noise to the counts in those cells.
    To control sensitivity, we normalize the trajectories so that each trajectory contributes at most 1 to the counts (the sensitivity is bounded to 1, see \autoref{app:method_grid}), and the privacy budget consumed is $\epsilon_1$.  We further partition dense cells (cells whose noisy counts are above a threshold) by a more fine-grained grid, but that step, by the postprocessing property of DP (see \autoref{subsec:grid}), consumes no privacy budget.
    
    \item \textbf{Markov Chain Models Learning.}
    \mymethod estimates the count of state transition to build the first- and the second-order Markov models (the states of the Markov models are the cells from the previous step).
    Similar to the previous step, we normalize the trajectory so that sensitivity is 1 (refer to \autoref{pf:transition_model_sensitivity}), and then add Laplace noise.
    The privacy budget consumed for the first- and the second-order Markov models are $\epsilon_2$ and $\epsilon_3$, respectively.
    
    \item \textbf{Trip Distribution Estimation.}
    \mymethod constructs an optimization problem to estimate the trip distribution.
    The data used in the optimization problem is gathered in the differentially private Markov models; thus the trip distribution estimation is a postprocessing operation, and no privacy budget is consumed.
    
    \item \textbf{Trajectories Generation}
    Trajectory generation uses data estimated in the previous components; it is also a postprocessing operation.
    Concretely, the information that we use to choose between the first- and second-order Markov models for next step prediction is the noisy counts in the first-order Markov model; thus, mixing the two Markov models for prediction is a postprocessing operation and does not consume extra privacy budget.
\end{itemize}

By the sequential composition property of DP as discussed in \autoref{subsec:composition}, \mymethod satisfies $\epsilon$-DP, where $\epsilon=\epsilon_1+\epsilon_2+\epsilon_3$.
\end{proof}

%% file: app_more.tex
\autoref{alg:norm-cut} illustrates the workflow of NormCut.
Suppose that we have a vector $(-5, 1, 7)$ with a negative value, the NormCut method takes the following steps to make the vector non-negative.
After executing \autoref{In2:rw1} to \autoref{In2:rw2}, the vector becomes $(-4, 0, 7)$.
Since it still has negative value, we then repeat \autoref{In2:rw1} to \autoref{In2:rw2} and obtain the non-negative vector $(0, 0, 3)$.

\section{More Details of Impact on Subgroups}
\label{subapp:subgroup}

\mypara{Details of~\cite{ganev2021robin}}
The authors divide the Adult dataset $R$ into subgroups according to the income attribute: The groups of records whose income exceeds 50K per year $R_{high}$ and the groups of records whose income is less than 50K per year $R_{low}$.
The synthetic dataset $S$ is also divided into two groups $S_{high}$ and $S_{low}$ following the same rule.
Two groups of evaluation are conducted to show the impact of DP data generation algorithms on the subgroups.
First, the authors compare the ratio $|R_{high}|/|R|$ and $|S_{high}|/|S|$ (also $|R_{low}|/|R|$ and $|S_{low}|/|S|$) to show that the DP data generation algorithms have impact on the ratio of subgroup.
Second, the authors train DP classifiers on the original data and non-DP classifiers on the synthetic data, then compare the prediction precision of these classifiers on subgroups to show how the DP data generation algorithms affect the data utility of subgroups for machine learning tasks. 
A similar evaluation is conducted on the Texas and Purchases datasets.
From their evaluation, they conclude that the DP data generation algorithms can affect both subgroup ratios and the data utility of subgroups.

\begin{figure*}[!htb]
    \centering
    
    \subfloat{\includegraphics[width=0.9\textwidth]{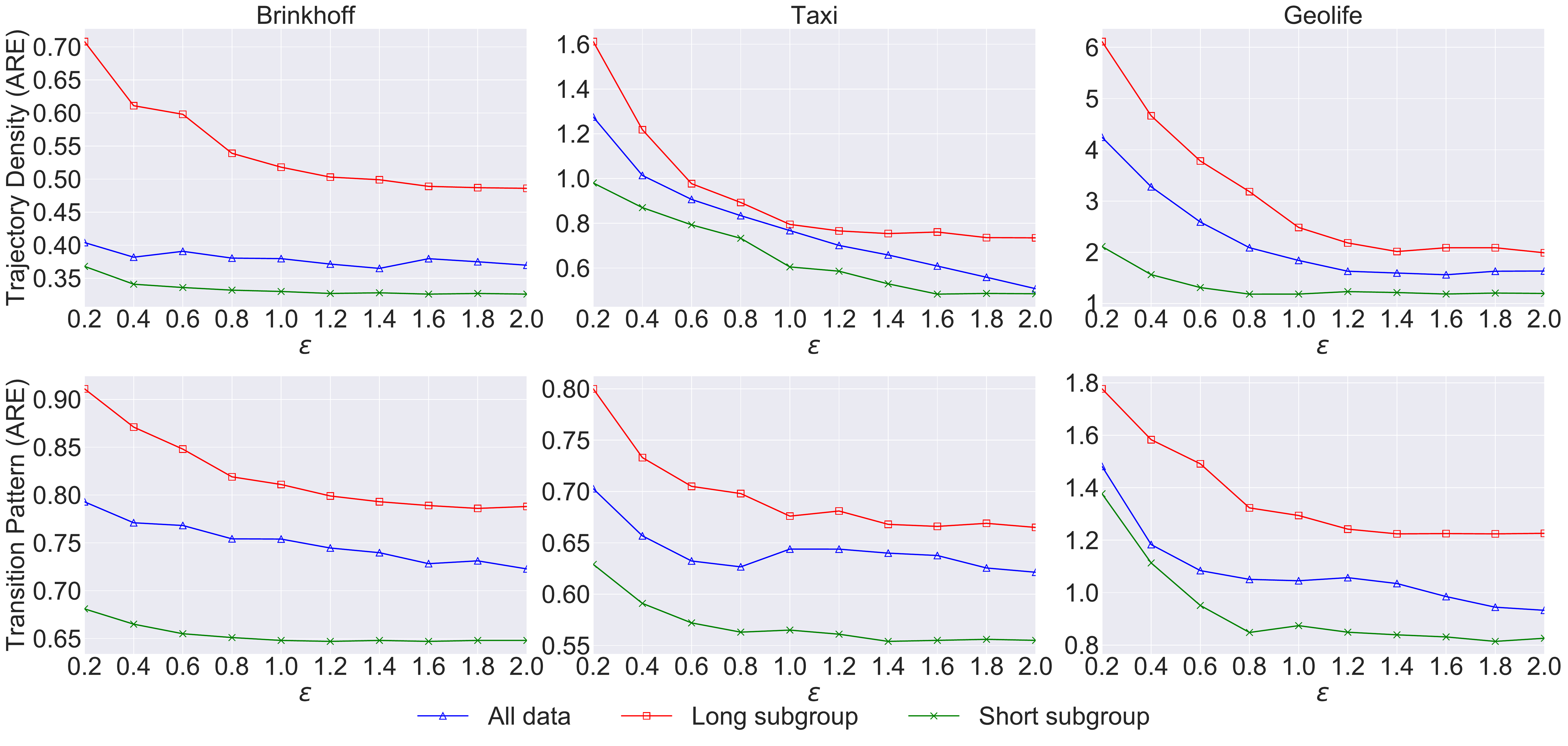}}

    \caption{Errors of the long subgroup and the short subgroup on the trajectory density and transition pattern metric for \adatrace.}
    \label{fig:ada_ratio_utility}
\end{figure*}

\begin{figure*}[!htb]
    \centering
    
    \subfloat{\includegraphics[width=0.9\textwidth]{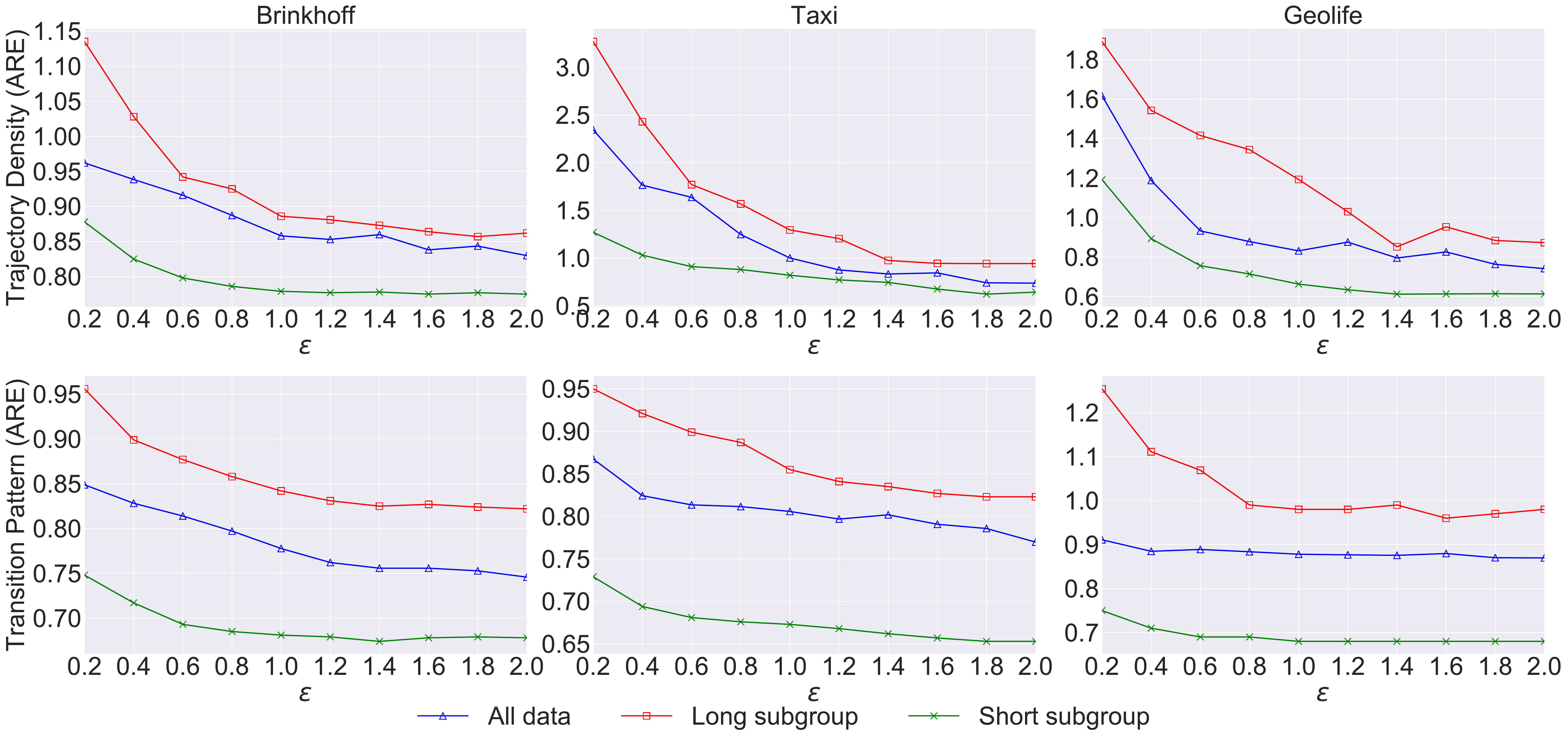}}

    \caption{Errors of the long subgroup and the short subgroup on the trajectory density and transition pattern metric for \dpt.}
    \label{fig:dpt_ratio_utility}
\end{figure*}

\mypara{Results for \adatrace and \dpt}
In Section 5.4, we conduct evaluations on the utility of subgroups and find that the short subgroup has better performance.
For \adatrace and \dpt, the same conclusion stands, which is shown in \autoref{fig:ada_ratio_utility} and \autoref{fig:dpt_ratio_utility}, respectively.

\section{Verifying Discretization Parameters}
\label{subapp:verify_discretization_parameters}

In this section, we empirically evaluate the effectiveness of our proposed parameter selection strategies for the discretization parameters in Section 5.1.
First, we use a $\levelonedipar \times \levelonedipar$ to divide the whole geographical space and the first-layer cells.
If a cell is passed through by many trajectories, we use a $\leveltwodipar \times \leveltwodipar$ grid to divide it again.

\mypara{Setup}
We conduct experiments on the Taxi dataset and four metrics.
We consider three different total privacy budgets: 0.2, 1.0, and 2.0.
Considering that $\leveltwodipar$ are different for different first-level cells, we introduce a variable $\psi$ that is easier to represent the granularity of the second-layer grid. $\psi$ is defined as $\sum_i\mathbbm{1}_{\leveltwodipar_i>1}$, the number of first-layer cells whose second-layer discretization parameter $\kappa$ is larger than $1$.
In other words, $\psi$ is the number of first-layer cells that are discretized by the second-layer grid. A larger $\psi$ means that more first-level cells are divided by the second-layer grids, indicating more fine-grained second-layer discretization.

\begin{figure*}[!tpb]
    \subfloat{\includegraphics[width=1\textwidth]{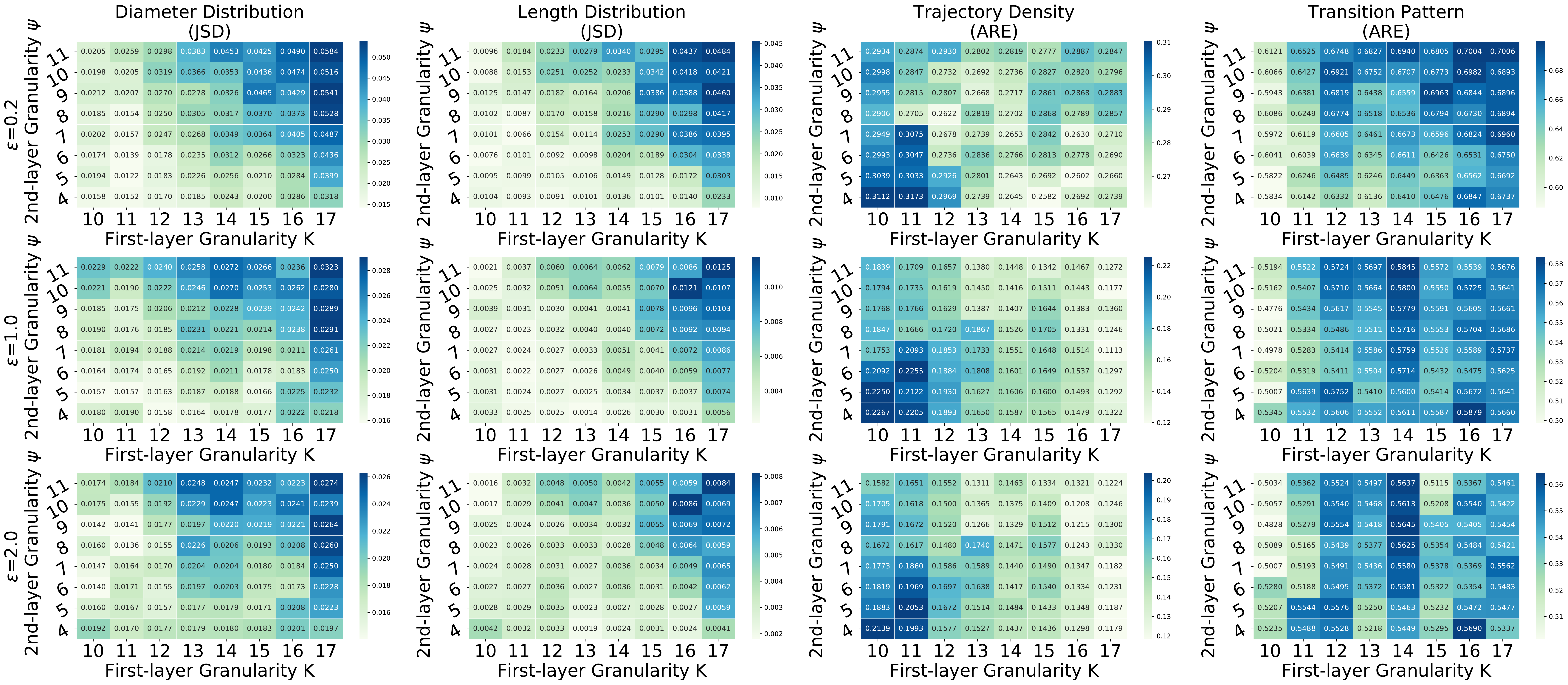}}

    \caption{Impact of grid granularity.
    In each subfig, the values stand for the errors of each metric, darker cells indicate larger errors.
    The $x$-axis and $y$-axis represent the first-layer granularity parameter $K$ and second-layer granularity parameter $\psi$, respectively.
    }
    \label{fig:grid_comparation}
\end{figure*}

\mypara{Observations}
\autoref{fig:grid_comparation} illustrates the experimental results.
In each subfig, the values stand for the errors of each metric, darker cells indicate larger errors.
The $x$-axis represents the first-layer granularity parameter $K$ and the $y$-axis represent $\psi$.
Errors under different $\psi$ can show the impact of different discretization granularities of the second layer.

The experimental results show that the diameter distribution and length distribution error increase when the granularity gets finer in both first-layer and second-layer discretization.
This is because the length and diameter are sensitive to noise in the first-order \markovmodel and second-order \markovmodel.
The scale of noise in the first-order and second-order \markovmodel increase with the square and cubic of the number of states, respectively.
On the other hand, for the trajectory density, the errors increase when the discretization granularity gets coarser.
This can be explained by the fact that the discrete trajectories obtained from coarse-grained discretization will lose much location information.
Therefore, the synthetic trajectories generated using these discrete sequences will not have precise density information.

These observations motivate us to choose parameters that reside in the center of each heatmap, which is consistent with our parameter selection strategy in Section 5.1.
Concretely, for the first-layer granularity, we have 
$\levelonedipar=(200000/1200)^\frac{1}{2} \approx 13$.
For the second-layer grid, we get the trajectory density $\levelonedensity$ for every first-layer cell and use $(\levelonedensity_{i} \times \levelonedipar \times pop / 20000)^{\frac{1}{2}}$ to calculate $\leveltwodipar$.
Here, by searching the Internet, we learn that Porto, the city where the Taxi dataset resides, has a population of 1,310,000.
After calculation, we have $7$ first-layer cells whose $\leveltwodipar$ are more than $1$.
That is, we set $\psi=7$.
We can see that $K=13$ and $\psi=7$ reside in the center of all heatmaps.

\begin{figure}[!tpb]
    \centering

    \subfloat{\includegraphics[width=0.5\textwidth]{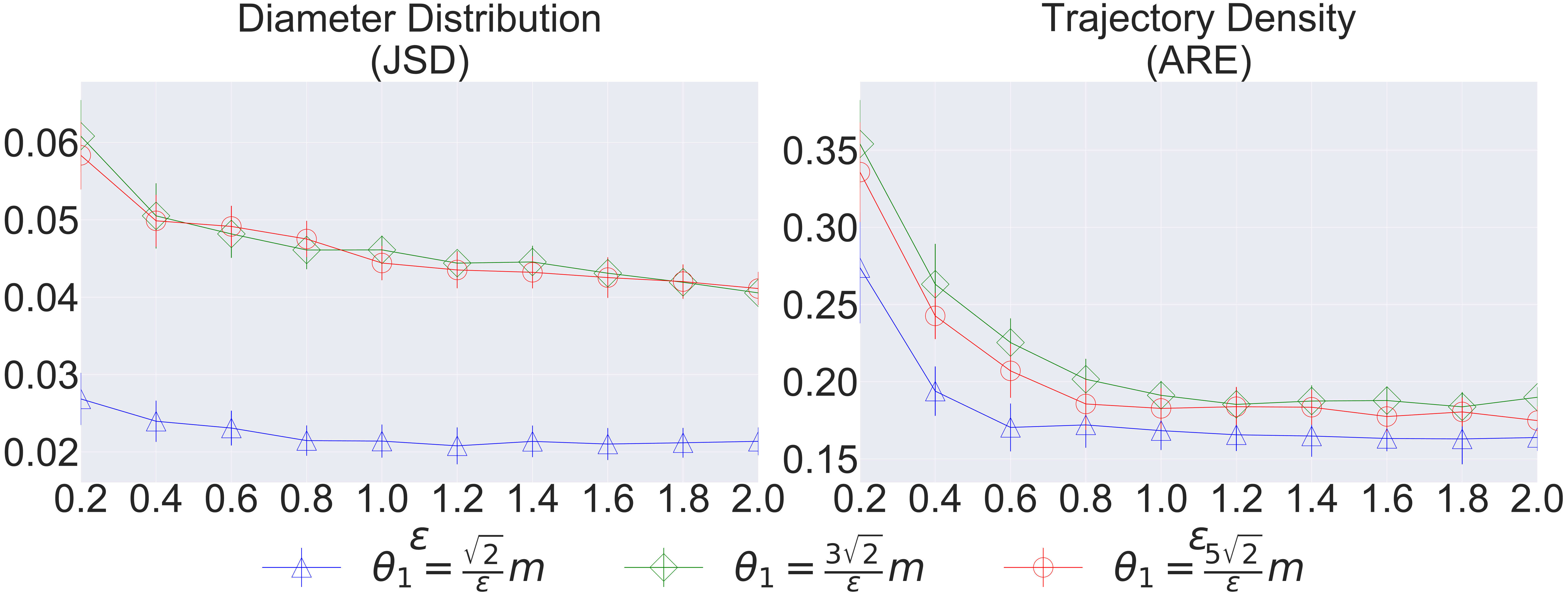}}
    \caption{Performance of different $\theta_1$.
    We use $\frac{\sqrt{2}}{\epsilon}\statenumber$ in \mymethod.
    }
    \label{fig:theta1}
\end{figure}

\begin{figure}[!tpb]
    \centering
    \subfloat{\includegraphics[width=0.5\textwidth]{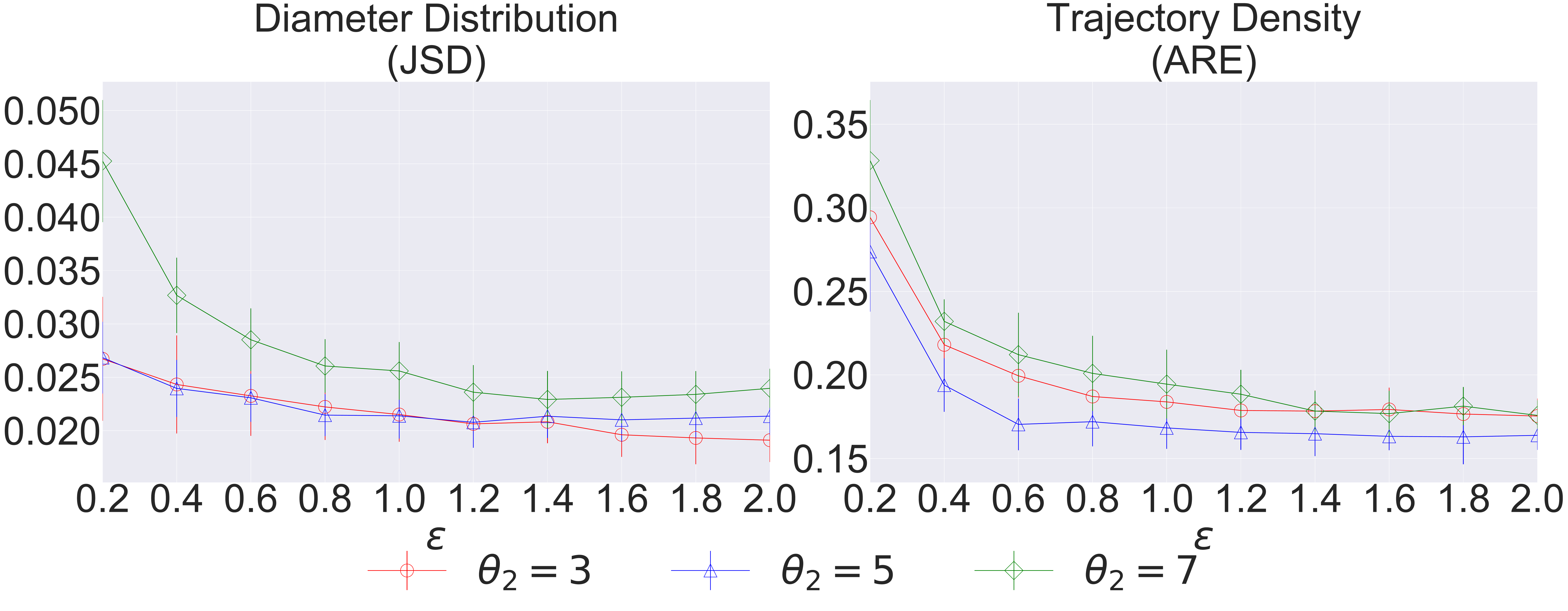}}
    \caption{Performance of different $\theta_2$.
    We use 5 in \mymethod.
    }
    \label{fig:theta2}
\end{figure}

\section{Verifying Model Selection Parameters}
\label{subapp:verify_model_selection_parameters}

In this section, we empirically validate the effectiveness of our proposed model selection parameter settings in Section 4.6.

\mypara{Setup}
We conduct experiments on the Taxi dataset and two metrics.
The conclusions for other datasets and metrics are consistent.
For both $\modelselectionpar_1$ and $\modelselectionpar_2$, we set three values in the experiments.
For $\modelselectionpar_1$, we conduct experiments on the value of $\frac{\sqrt{2}}{\epsilon}\statenumber$, $\frac{3\sqrt{2}}{\epsilon}\statenumber$
, and $\frac{5\sqrt{2}}{\epsilon}\statenumber$, among which $\frac{\sqrt{2}}{\epsilon}\statenumber$ is used in \mymethod.
For $\modelselectionpar_2$, we conduct experiments on the value of 3, 5, and 7, among which 5 is used in \mymethod.
We conduct all experiments with the privacy budget varying from 0.2 to 2.0.

\mypara{Observations}
In general, the experimental results in \autoref{fig:theta1} and \autoref{fig:theta2} show that the parameters used in \mymethod achieve the best performance.
Concretely, for $\modelselectionpar_1$, we can observe that $\modelselectionpar_1=\frac{\sqrt{2}}{\epsilon}\statenumber$ has the smallest error on both metrics. 
This result can be explained by the fact that if $\modelselectionpar_1$ is very large, most of the states will choose the first-order Markov chain model, which will lose much helpful information in the second-order Markov model.
For $\modelselectionpar_2$, we find that $\modelselectionpar_2=5$ has the best performance in most cases.
The reason is that we reduce the noise in the algorithm by selecting the first-order Markov chain model for states for which the first-order model has high confidence in the next step prediction.

\begin{figure*}[!htb]
    \centering
    
    \subfloat{\includegraphics[width=1\textwidth]{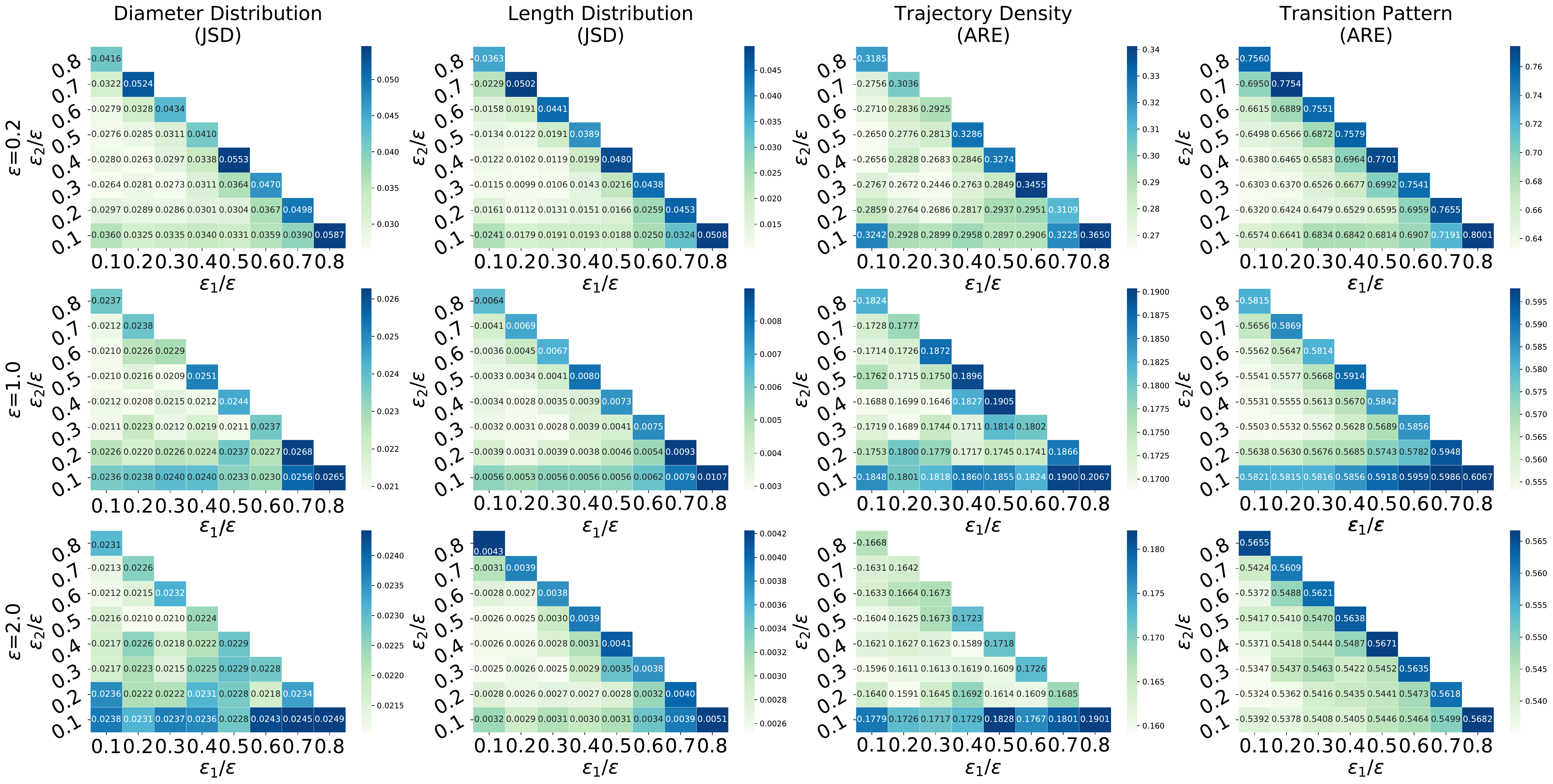}}

    \caption{Impact of privacy budget allocation.
    In each subfig, the values stand for the errors of each metric, darker cells indicate larger errors.
    The $x$-axis and $y$-axis stands for the ratio of $\epsilon_1$ and $\epsilon_2$ to $\epsilon$.
    Note that $\epsilon_3=\epsilon-\epsilon_1-\epsilon_2$.}
    \label{fig:partition}
\end{figure*}

\section{Privacy Budget Allocation}
\label{subapp:impact_budget_allocation}

In this section, we evaluate the impact of different privacy budget allocation strategies and identify a practical allocation strategy in real-world.

\mypara{Setup}
Recall that, in \mymethod, the total privacy budget $\epsilon$ is divided into three parts: Space discretization ($\epsilon_1$), the first-order \markovmodel learning ($\epsilon_2$), and the second-order \markovmodel learning ($\epsilon_3$).
We vary the ratio of $\epsilon_1$ and $\epsilon_2$ to $\epsilon$ from 0.1 to 0.8 with step size 0.1.
Note that $\epsilon_3$ is automatically calculated as $\epsilon-\epsilon_1-\epsilon_2$.
We conduct experiments on three total privacy budgets ($\epsilon$): 0.2, 1.0, 2.0.

\mypara{Observations}
\autoref{fig:partition} shows algorithm performance under different $\epsilon_{1}$, $\epsilon_{2}$ and $\epsilon_{3}$.
For all heatmaps, results in the leftmost column, bottom row, and diagonal correspond to the settings where $\epsilon_{1}$, $\epsilon_{2}$, and $\epsilon_{3}$ have the smallest values, respectively.
We observe that \mymethod performs differently in these settings.
To analyze the impact of small $\epsilon_1$, we compare the errors in the leftmost column with the bottom row and diagonal.
In general, the error caused by a small $\epsilon_1$ is far smaller than the error caused by a small $\epsilon_2$ or $\epsilon_3$.
To analyze the impact of small $\epsilon_3$, we compare the errors in the diagonal with the other two conditions.
Contrary to small $\epsilon_1$, the error caused by a small $\epsilon_3$ is the largest in all cases.
By analyzing the bottom row, we conclude that a small $\epsilon_2$ will cause an error between the two conditions above.
In addition, we notice that in some cases (such as $\epsilon=2.0$, Diameter distribution), the error caused by a small $\epsilon_2$ is larger than that caused by a small $\epsilon_3$.
This can be explained by the fact that $\marquery_1$ is not only used in the generation but also the trip distribution estimation.
Thus, a small $\epsilon_2$ may also cause a bad result.

\mypara{Guideline}
The above observations and analysis provide us a guideline to choose the privacy budget allocation.
That is, we should allocate more privacy budget to $\epsilon_2$ and $\epsilon_3$, and allocate less to $\epsilon_1$.
In our experiments, we set $\epsilon_1$, $\epsilon_2$, and $\epsilon_3$ as $0.2 \cdot \epsilon$, $0.4 \cdot \epsilon$, and $0.4 \cdot \epsilon$ respectively.
The experimental results in \autoref{fig:partition} show that this privacy budget allocation strategy can obtain promising performance.

%% file: main.bbl
\begin{thebibliography}{10}

\bibitem{abadi2016deep}
Martin Abadi, Andy Chu, Ian Goodfellow, H~Brendan McMahan, Ilya Mironov, Kunal
  Talwar, and Li~Zhang.
\newblock {Deep Learning with Differential Privacy}.
\newblock In {\em Proceedings of the 2016 ACM SIGSAC conference on computer and
  communications security}, pages 308--318, 2016.

\bibitem{abay2018privacy}
Nazmiye~Ceren Abay, Yan Zhou, Murat Kantarcioglu, Bhavani Thuraisingham, and
  Latanya Sweeney.
\newblock {Privacy Preserving Synthetic Data Release Using Deep Learning}.
\newblock In {\em Joint European Conference on Machine Learning and Knowledge
  Discovery in Databases}, pages 510--526. Springer, 2018.

\bibitem{acs2014case}
Gergely Acs and Claude Castelluccia.
\newblock {A Case Study: Privacy Preserving Release of Spatio-temporal Density
  in Paris}.
\newblock In {\em Proceedings of the 20th ACM SIGKDD international conference
  on Knowledge discovery and data mining}, pages 1679--1688, 2014.

\bibitem{atluri2018spatio}
Gowtham Atluri, Anuj Karpatne, and Vipin Kumar.
\newblock {Spatio-temporal Data Mining: a Survey of Problems and Methods}.
\newblock {\em ACM Computing Surveys (CSUR)}, 51(4):1--41, 2018.

\bibitem{bashir2007object}
Faisal~I Bashir, Ashfaq~A Khokhar, and Dan Schonfeld.
\newblock {Object Trajectory-based Activity Classification and Recognition
  Using Hidden Markov Models}.
\newblock {\em IEEE transactions on Image Processing}, 16(7):1912--1919, 2007.

\bibitem{beaulieu2019privacy}
Brett~K Beaulieu-Jones, Zhiwei~Steven Wu, Chris Williams, Ran Lee, Sanjeev~P
  Bhavnani, James~Brian Byrd, and Casey~S Greene.
\newblock {Privacy-preserving Generative Deep Neural Networks Support Clinical
  Data Sharing}.
\newblock {\em Circulation: Cardiovascular Quality and Outcomes},
  12(7):e005122, 2019.

\bibitem{bindschaedler2016synthesizing}
Vincent Bindschaedler and Reza Shokri.
\newblock {Synthesizing Plausible Privacy-preserving Location Traces}.
\newblock In {\em 2016 IEEE Symposium on Security and Privacy (SP)}, pages
  546--563. IEEE, 2016.

\bibitem{bindschaedler2017plausible}
Vincent Bindschaedler, Reza Shokri, and Carl~A Gunter.
\newblock {Plausible Deniability for Privacy-Preserving Data Synthesis}.
\newblock {\em Proceedings of the VLDB Endowment}, 10(5), 2017.

\bibitem{brinkhoff2002framework}
Thomas Brinkhoff.
\newblock {A Framework for Generating Network-based Moving Objects}.
\newblock {\em GeoInformatica}, 6(2):153--180, 2002.

\bibitem{CensusBureau}
United States~Census Bureau.
\newblock
  \url{https://www.census.gov/programs-surveys/decennial-census/decade/2020/planning-management/process/disclosure-avoidance/differential-privacy.html}.

\bibitem{chang2018revealing}
Shan Chang, Chao Li, Hongzi Zhu, Ting Lu, and Qiang Li.
\newblock {Revealing Privacy Vulnerabilities of Anonymous Trajectories}.
\newblock {\em IEEE Transactions on Vehicular Technology}, 67(12):12061--12071,
  2018.

\bibitem{CZWBHZ21}
Min Chen, Zhikun Zhang, Tianhao Wang, Michael Backes, Mathias Humbert, and Yang
  Zhang.
\newblock {When Machine Unlearning Jeopardize Privacy}.
\newblock In {\em {ACM CCS}}, 2021.

\bibitem{chen2012differentially}
Rui Chen, Gergely Acs, and Claude Castelluccia.
\newblock {Differentially Private Sequential Data Publication Via
  Variable-length N-grams}.
\newblock In {\em Proceedings of the 2012 ACM conference on Computer and
  communications security}, pages 638--649, 2012.

\bibitem{chen2015differentially}
Rui Chen, Qian Xiao, Yu~Zhang, and Jianliang Xu.
\newblock {Differentially Private High-dimensional Data Publication via
  Sampling-based Inference}.
\newblock In {\em Proceedings of the 21th ACM SIGKDD international conference
  on knowledge discovery and data mining}, pages 129--138, 2015.

\bibitem{cheng2016improved}
Yihang Cheng, Yuanyuan Qiao, and Jie Yang.
\newblock {An Improved Markov Method for Prediction of User Mobility}.
\newblock In {\em 2016 12th International Conference on Network and Service
  Management (CNSM)}, pages 394--399. IEEE, 2016.

\bibitem{cormode2012differentially}
Graham Cormode, Cecilia Procopiuc, Divesh Srivastava, Entong Shen, and Ting Yu.
\newblock {Differentially Private Spatial Decompositions}.
\newblock In {\em 2012 IEEE 28th International Conference on Data Engineering},
  pages 20--31. IEEE, 2012.

\bibitem{DZBLJCC21}
Linkang Du, Zhikun Zhang, Shaojie Bai, Changchang Liu, Shouling Ji, Peng Cheng,
  and Jiming Chen.
\newblock {AHEAD: Adaptive Hierarchical Decomposition for Range Query under
  Local Differential Privacy}.
\newblock In {\em {ACM CCS}}, 2021.

\bibitem{dwork2006calibrating}
Cynthia Dwork, Frank McSherry, Kobbi Nissim, and Adam Smith.
\newblock {Calibrating Noise to Sensitivity in Private data Analysis}.
\newblock In {\em Theory of cryptography conference}, pages 265--284. Springer,
  2006.

\bibitem{fine1998hierarchical}
Shai Fine, Yoram Singer, and Naftali Tishby.
\newblock {The Hierarchical Hidden Markov Model: Analysis and Applications}.
\newblock {\em Machine learning}, 32(1):41--62, 1998.

\bibitem{frigerio2019differentially}
Lorenzo Frigerio, Anderson~Santana de~Oliveira, Laurent Gomez, and Patrick
  Duverger.
\newblock {Differentially Private Generative Adversarial Networks for Time
  Series, Continuous, and Discrete Open Data}.
\newblock In {\em IFIP International Conference on ICT Systems Security and
  Privacy Protection}, pages 151--164. Springer, 2019.

\bibitem{gaboardi2014dual}
Marco Gaboardi, Emilio Jes{\'u}s~Gallego Arias, Justin Hsu, Aaron Roth, and
  Zhiwei~Steven Wu.
\newblock {Dual Query: Practical Private Query Release for High Dimensional
  Data}.
\newblock In {\em International Conference on Machine Learning}, pages
  1170--1178. PMLR, 2014.

\bibitem{gambs2012next}
S{\'e}bastien Gambs, Marc-Olivier Killijian, and Miguel~N{\'u}{\~n}ez del
  Prado~Cortez.
\newblock {Next Place Prediction Using Mobility Markov Chains}.
\newblock In {\em Proceedings of the first workshop on measurement, privacy,
  and mobility}, pages 1--6, 2012.

\bibitem{ganev2021robin}
Georgi Ganev, Bristena Oprisanu, and Emiliano De~Cristofaro.
\newblock {Robin Hood and Matthew Effects--Differential Privacy Has Disparate
  Impact on Synthetic Data}.
\newblock {\em arXiv preprint arXiv:2109.11429}, 2021.

\bibitem{goh2012online}
Chong~Yang Goh, Justin Dauwels, Nikola Mitrovic, Muhammad~Tayyab Asif, Ali
  Oran, and Patrick Jaillet.
\newblock {Online Map-matching Based on Hidden Markov Model for Real-time
  Traffic Sensing Applications}.
\newblock In {\em 2012 15th International IEEE Conference on Intelligent
  Transportation Systems}, pages 776--781. IEEE, 2012.

\bibitem{gursoy2018utility}
Mehmet~Emre Gursoy, Ling Liu, Stacey Truex, Lei Yu, and Wenqi Wei.
\newblock {Utility-aware Synthesis of Differentially Private and
  Attack-resilient Location Traces}.
\newblock In {\em Proceedings of the 2018 ACM SIGSAC Conference on Computer and
  Communications Security}, pages 196--211, 2018.

\bibitem{adatracecode}
Mehmet~Emre Gursoy, Ling Liu, Stacey Truex, Lei Yu, and Wenqi Wei.
\newblock {Adatrace-Github}.
\newblock \url{https://github.com/git-disl/AdaTrace}, 2019.

\bibitem{hardt2012simple}
Moritz Hardt, Katrina Ligett, and Frank McSherry.
\newblock {A Simple and Practical Algorithm for Differentially Private Data
  Release}.
\newblock In {\em Proceedings of the 25th International Conference on Neural
  Information Processing Systems-Volume 2}, pages 2339--2347, 2012.

\bibitem{he2015dpt}
Xi~He, Graham Cormode, Ashwin Machanavajjhala, Cecilia~M Procopiuc, and Divesh
  Srivastava.
\newblock {DPT: Differentially Private Trajectory Synthesis using Hierarchical
  Reference Systems}.
\newblock {\em Proceedings of the VLDB Endowment}, 8(11):1154--1165, 2015.

\bibitem{iwata2019neural}
Tomoharu Iwata and Hitoshi Shimizu.
\newblock {Neural Collective Graphical Models for Estimating Spatio-temporal
  Population Flow from Aggregated Data}.
\newblock In {\em Proceedings of the AAAI Conference on Artificial
  Intelligence}, volume~33, pages 3935--3942, 2019.

\bibitem{jiang2013publishing}
Kaifeng Jiang, Dongxu Shao, St{\'e}phane Bressan, Thomas Kister, and Kian-Lee
  Tan.
\newblock {Publishing Trajectories with Differential Privacy Guarantees}.
\newblock In {\em Proceedings of the 25th International Conference on
  Scientific and Statistical Database Management}, pages 1--12, 2013.

\bibitem{juang1991hidden}
Biing~Hwang Juang and Laurence~R Rabiner.
\newblock {Hidden Markov Models for Speech Recognition}.
\newblock {\em Technometrics}, 33(3):251--272, 1991.

\bibitem{karimzadeh2018pedestrians}
Mostafa Karimzadeh, Zhongliang Zhao, Florian Gerber, and Torsten Braun.
\newblock {Pedestrians Complex Behavior Understanding and Prediction with
  Hybrid Markov Chain}.
\newblock In {\em 2018 14th International Conference on Wireless and Mobile
  Computing, Networking and Communications (WiMob)}, pages 200--207. IEEE,
  2018.

\bibitem{li2017achieving}
Meng Li, Liehuang Zhu, Zijian Zhang, and Rixin Xu.
\newblock {Achieving Differential Privacy of Trajectory Data Publishing in
  Participatory Sensing}.
\newblock {\em Information Sciences}, 400:1--13, 2017.

\bibitem{lin1991divergence}
Jianhua Lin.
\newblock {Divergence Measures Based on the Shannon Entropy}.
\newblock {\em IEEE Transactions on Information theory}, 37(1):145--151, 1991.

\bibitem{lv2016big}
Qiujian Lv, Yuanyuan Qiao, Nirwan Ansari, Jun Liu, and Jie Yang.
\newblock {Big Data Driven Hidden Markov Model Based Individual Mobility
  Prediction at Points of Interest}.
\newblock {\em IEEE Transactions on Vehicular Technology}, 66(6):5204--5216,
  2016.

\bibitem{mckenna2019graphical}
Ryan McKenna, Daniel Sheldon, and Gerome Miklau.
\newblock {Graphical-model Based Estimation and Inference for Differential
  Privacy}.
\newblock In {\em International Conference on Machine Learning}, pages
  4435--4444. PMLR, 2019.

\bibitem{moreira2013predicting}
Luis Moreira-Matias, Joao Gama, Michel Ferreira, Joao Mendes-Moreira, and Luis
  Damas.
\newblock {Predicting Taxi--passenger Demand Using Streaming Data}.
\newblock {\em IEEE Transactions on Intelligent Transportation Systems},
  14(3):1393--1402, 2013.

\bibitem{morwal2012named}
Sudha Morwal, Nusrat Jahan, and Deepti Chopra.
\newblock {Named Entity Recognition Using Hidden Markov Model (HMM)}.
\newblock {\em International Journal on Natural Language Computing (IJNLC)},
  1(4):15--23, 2012.

\bibitem{pyrgelis2017knock}
Apostolos Pyrgelis, Carmela Troncoso, and Emiliano De~Cristofaro.
\newblock {Knock Knock, Who's There? Membership Inference on Aggregate Location
  Data}.
\newblock {\em arXiv preprint arXiv:1708.06145}, 2017.

\bibitem{dpgrid2013}
Wahbeh Qardaji, Weining Yang, and Ninghui Li.
\newblock {Differentially Private Grids for Geospatial Data}.
\newblock In {\em 2013 IEEE 29th international conference on data engineering
  (ICDE)}, pages 757--768. IEEE, 2013.

\bibitem{song2006evaluating}
Libo Song, David Kotz, Ravi Jain, and Xiaoning He.
\newblock {Evaluating Next-cell Predictors with Extensive Wi-Fi Mobility Data}.
\newblock {\em IEEE transactions on mobile computing}, 5(12):1633--1649, 2006.

\bibitem{stadler2021synthetic}
Theresa Stadler, Bristena Oprisanu, and Carmela Troncoso.
\newblock {Synthetic Data--Anonymisation Groundhog Day}.
\newblock {\em arXiv preprint arXiv:2011.07018}, 2021.

\bibitem{tantipongpipat2019differentially}
Uthaipon Tantipongpipat, Chris Waites, Digvijay Boob, Amaresh~Ankit Siva, and
  Rachel Cummings.
\newblock {Differentially Private Mixed-type Data Generation for Unsupervised
  Learning}.
\newblock {\em arXiv preprint arXiv:1912.03250}, 2019.

\bibitem{tokuda2013speech}
Keiichi Tokuda, Yoshihiko Nankaku, Tomoki Toda, Heiga Zen, Junichi Yamagishi,
  and Keiichiro Oura.
\newblock {Speech Synthesis Based on Hidden Markov Models}.
\newblock {\em Proceedings of the IEEE}, 101(5):1234--1252, 2013.

\bibitem{vietri2020new}
Giuseppe Vietri, Grace Tian, Mark Bun, Thomas Steinke, and Steven Wu.
\newblock {New Oracle-efficient Algorithms for Private Synthetic Data Release}.
\newblock In {\em International Conference on Machine Learning}, pages
  9765--9774. PMLR, 2020.

\bibitem{wang2020deep}
Senzhang Wang, Jiannong Cao, and Philip Yu.
\newblock {Deep Learning for Spatio-temporal Data Mining: a Survey}.
\newblock {\em IEEE transactions on knowledge and data engineering}, 2020.

\bibitem{WCZSCLLJ21}
Tianhao Wang, Joann~Qiongna Chen, Zhikun Zhang, Dong Su, Yueqiang Cheng, Zhou
  Li, Ninghui Li, and Somesh Jha.
\newblock {Continuous Release of Data Streams under both Centralized and Local
  Differential Privacy}.
\newblock In {\em {ACM CCS}}, 2021.

\bibitem{wang2020locally}
Tianhao Wang, Milan Lopuhaa-Zwakenberg, Zitao Li, Boris Skoric, and Ninghui Li.
\newblock {Locally Differentially Private Frequency Estimation with
  Consistency}.
\newblock In {\em NDSS'20: Proceedings of the NDSS Symposium}, 2020.

\bibitem{wei2019differential}
Jianhao Wei, Yaping Lin, Xin Yao, and Voundi Koe~Arthur Sandor.
\newblock {Differential Privacy-based Trajectory Community Recommendation in
  Social Network}.
\newblock {\em Journal of Parallel and Distributed Computing}, 133:136--148,
  2019.

\bibitem{DPTcode}
H.~Xi and M.~Ashwin.
\newblock {Differentially Private Trajectory Synthesis}.
\newblock \url{https://users.cs.duke.edu/~hexi88/project_dpt/index.html}, 2015.

\bibitem{xiao2010differentially}
Yonghui Xiao, Li~Xiong, and Chun Yuan.
\newblock {Differentially Private Data Release Through Multidimensional
  Partitioning}.
\newblock In {\em Workshop on Secure Data Management}, pages 150--168.
  Springer, 2010.

\bibitem{xu2018dp}
Changqiao Xu, Liang Zhu, Yang Liu, Jianfen Guan, and Shui Yu.
\newblock {Dp-ltod: Differential Privacy Latent Trajectory Community
  Discovering Services over Location-based Social Networks}.
\newblock {\em IEEE Transactions on Services Computing}, 2018.

\bibitem{zhang2017privbayes}
Jun Zhang, Graham Cormode, Cecilia~M Procopiuc, Divesh Srivastava, and Xiaokui
  Xiao.
\newblock {Privbayes: Private Data Release Via Bayesian Networks}.
\newblock {\em ACM Transactions on Database Systems (TODS)}, 42(4):1--41, 2017.

\bibitem{zhang2018differentially}
Xinyang Zhang, Shouling Ji, and Ting Wang.
\newblock {Differentially Private Releasing Via Deep Generative Model
  (technical report)}.
\newblock {\em arXiv preprint arXiv:1801.01594}, 2018.

\bibitem{ZWLHC18}
Zhikun Zhang, Tianhao Wang, Ninghui Li, Shibo He, and Jiming Chen.
\newblock {CALM: Consistent Adaptive Local Marginal for Marginal Release under
  Local Differential Privacy}.
\newblock In {\em {ACM CCS}}, 2018.

\bibitem{zhang2022privsyn}
Zhikun Zhang, Tianhao Wang, Ninghui Li, Jean Honorio, Michael Backes, Shibo He,
  Jiming Chen, and Yang Zhang.
\newblock {PrivSyn: Differentially Private Data Synthesis}.
\newblock {\em USENIX Security}, 2022.

\bibitem{zheng2009mining}
Yu~Zheng, Lizhu Zhang, Xing Xie, and Wei-Ying Ma.
\newblock {Mining Interesting Locations and Travel Sequences from Gps
  Trajectories}.
\newblock In {\em Proceedings of the 18th international conference on World
  wide web}, pages 791--800, 2009.

\end{thebibliography}
